\documentclass[aps,prx,twocolumn,showpacs,superscriptaddress,floatfix
,nofootinbib]{revtex4-1}
\usepackage{amsmath}
\usepackage{graphicx}% Include figure files
\usepackage{graphicx}
\usepackage{dcolumn}% Align table columns on decimal point
\usepackage{bm}% bold math
\usepackage{hyperref}% add hypertext capabilities
\usepackage{ulem}
\usepackage{xcolor}
\usepackage[utf8]{inputenc}
\usepackage{tabularx}
\usepackage{color}
\usepackage{cellspace}
\usepackage{lipsum, babel}
\usepackage{lipsum}% http://ctan.org/pkg/lips
\usepackage{multirow}
\usepackage{booktabs} 
\definecolor{customviolet}{RGB}{148,0,211}
\usepackage{orcidlink}
\newcommand{\be}{\begin{equation}}
\newcommand{\ee}{\end{equation}}
\newcommand{\bea}{\begin{eqnarray}}
\newcommand{\eea}{\end{eqnarray}}

\begin{document}

\preprint{APS/123-QED}

\title{Feasibility of dark matter admixed neutron star based on recent observational constraints}

\author{Prashant Thakur \orcidlink{0000-0003-4189-6176}}
\email{p20190072@goa.bits-pilani.ac.in}
\affiliation{Department of Physics, BITS-Pilani, K. K. Birla Goa Campus, Goa 403726, India}
\affiliation{CFisUC, Department of Physics, University of Coimbra, P-3004 - 516  Coimbra, Portugal}
\author{Tuhin Malik \orcidlink{0000-0003-2633-5821}}
\email{tuhin.malik@uc.pt}
\affiliation{CFisUC, Department of Physics, University of Coimbra, P-3004 - 516  Coimbra, Portugal}
\author{Arpan Das \orcidlink{0000-0002-2630-8914}}
\email{arpan.das@pilani.bits-pilani.ac.in}
\affiliation{Department of Physics,  Birla Institute of Technology and Science Pilani, Pilani Campus, Pilani,  Rajasthan-333031, India} 

\author{T. K. Jha \orcidlink{0000-0002-9334-240X}}
\email{tkjha@goa.bits-pilani.ac.in}
\affiliation{Department of Physics, BITS-Pilani, K. K. Birla Goa Campus, Goa 403726, India}
\author{B. K. Sharma \orcidlink{0000-0003-0543-1582}}
\email{bk\_sharma@cb.amrita.edu}
\affiliation{Department of Sciences, Amrita School of Physical Sciences, Amrita Vishwa Vidyapeetham, Coimbatore 641112, India}
\author{Constança Providência \orcidlink{0000-0001-6464-8023}}
\email{cp@uc.pt}
\affiliation{CFisUC, Department of Physics, University of Coimbra, P-3004 - 516  Coimbra, Portugal}

\date{\today}
\begin{abstract}
The equation of state (EOS) for neutron stars is modeled using the Relativistic Mean Field (RMF) approach with a mesonic nonlinear (NL) interaction, a modified sigma cut potential (NL-$\sigma$ cut), and the influences of dark matter in the NL (NL DM). Using a Bayesian analysis framework, we evaluate the plausibility and impact of each scenario. Experimental constraints on the general properties of finite nuclei and heavy ion collisions, along with astrophysical observational data on neutron star radii and tidal deformation, have been taken into account. It was shown that all models, including the PREX-II data, were less favored, indicating that this experimental data seemed to be in tension with the other constraints included in the inference procedure, and were incompatible with chiral effective field theoretical calculations of pure neutron matter. Considering the models with no PREX-II constraints, we find the model NL-$\sigma$ cut with the largest Bayes evidence, indicating that the constraints considered favor the stiffening of the EOS at large densities. Conversely, the neutron star with a dark matter component is the least favorable case in light of recent observational constraints, among different scenarios considered here. The $f$ and $p$ modes were calculated within the Cowling approximation, and it can be seen that $f$ modes are sensitive to the EOS. An analysis of the slopes of the mass-radius curves and $f$-mode mass curves has indicated that these quantities may help distinguish the different scenarios.We also analyzed the impact of new PSR J0437-4715 measurements on neutron star mass-radius estimates, noting a $\sim$ 0.2 km reduction in the 90\% CI upper boundary across all models and a significant Bayes evidence decrease, indicating potential conflicts with previous data or the necessity for more adaptable models.
\end{abstract}

\maketitle
\section{Introduction}
The equation of state (EOS) of neutron stars is pivotal in determining their internal structure and observable characteristics, such as mass and radius. Various microscopic many-body approaches and phenomenological models are employed to construct the EOS, taking into account the interactions of nucleons, hyperons, and other exotic particles \cite{Gal:2016boi,Curceanu:2019uph,Tolos:2020aln}. The EOS models are validated against both terrestrial laboratory data and astrophysical observations \cite{Malik:2022jqc, Dutra:2014qga,Oertel:2016bki,Burgio:2021vgk}. Furthermore, the detection of gravitational waves from neutron star mergers provides additional constraints, with postmerger emissions offering insights into the EOS by correlating specific spectral features with the star's compactness \cite{Takami:2014zpa}. {Recent studies have shown that the onset of phase transition from hadronic matter to quark matter can significantly affect the dynamics and gravitational wave signals of binary neutron star mergers \cite{Weih:2019xvw,Haque:2022dsc}. One area of research that has recently attracted considerable interest is altered gravity theories, with the study of neutron stars within the framework of modified gravity theories being a lively field of research, offering an opportunity to compare the predictions of these theories with observational data \cite{Nobleson:2023wca,Alam:2023grx,Cooney:2009rr,Yazadjiev:2018xxk}.} The combined efforts from both theoretical and observational studies are essential for enhancing our understanding of the extreme conditions within neutron stars. Recent studies have extended this understanding by incorporating the effects of dark matter (DM) within neutron stars \cite{Ellis:2018bkr,Das:2018frc,Nelson:2018xtr,Das:2020ecp,Karkevandi:2021ygv,Lenzi:2022ypb,Giangrandi:2022wht,Rutherford:2022xeb, Routaray:2022utr,Shakeri:2022dwg,Singh:2022wvw, Thakur:2023aqm,Sagun:2023rzp,Diedrichs:2023trk,Cronin:2023xzc,Diedrichs:2023trk,Flores:2024hts, Scordino:2024ehe, Barbat:2024yvi}. Dark matter, interacting with neutrons through mechanisms such as Higgs boson exchange, significantly alters the EOS, potentially leading to observable changes in neutron star characteristics. For instance, DM can {affects} the mass-to-radius relation \cite{Das:2020vng, Das:2021yny, Das:2021dru, Sen:2021wev, Guha:2021njn, Lourenco:2022fmf, Sen:2022pfr, Shirke:2023ktu, Shirke:2024ymc, Panotopoulos:2017idn}, and the cooling rates of neutron stars \cite{Kouvaris:2007ay,Giangrandi:2024qdb}. Studies show that neutron stars containing a mix of baryonic and dark matter can exhibit different mass-radius relations compared to those composed solely of neutron matter, influencing gravitational redshift and potentially explaining observations inconsistent with normal neutron stars \cite{Rezaei:2016zje}.

Numerous candidates for dark matter particles, including bosonic dark matter, axions, sterile neutrinos, and various WIMPs, are discussed in \cite{Bertone:2010zza,Bauer:2017qwy,Calmet:2020pub}. The nature of dark matter and its properties, such as self-interaction and coupling with standard model particles, have been explored for their impact on neutron star dynamics \cite{Ellis:2018bkr,Panotopoulos:2017idn,Diedrichs:2023trk,Leung:2022wcf,Narain:2006kx}. Two commonly researched methods for examining neutron stars mixed with dark matter are found in existing literature. One approach considers non-gravitational interactions, using mechanisms like the Higgs portal \cite{Dutra:2022mxl, Lenzi:2022ypb, Hong:2024sey, Das:2018frc, Flores:2024hts, Das:2020vng, Sen:2021wev}. The other approach considers only gravitational interactions, treated as a two-fluid system \cite{Collier:2022cpr, Miao:2022rqj,Emma:2022xjs, Hong:2024sey, Karkevandi:2021ygv,Ruter:2023uzc,Liu:2023ecz,Ivanytskyi:2019wxd,Buras-Stubbs:2024don,Rutherford:2022xeb}. Another method involves the neutron decay into dark matter particles that accounts for the neutron decay anomaly \cite{Husain:2022bxl,Bastero-Gil:2024kjo,PhysRevLett.121.061801,Shirke:2023ktu,Motta:2018bil,Shirke:2024ymc}. This can be addressed using either a single-fluid method or a two-fluid method. 
Discrepancies in neutron decay lifetimes measured via bottle and beam experiments suggest more decayed neutrons than produced protons, possibly due to decay into nearly degenerate dark fermions.

To understand the mixed dark matter scenario in neutron stars (NS), knowledge of the EOS for both baryonic and dark matter is crucial. The dense matter EOS can be described by relativistic and nonrelativistic models. Nonrelativistic models describe nucleons within finite nuclei well but fail with infinite dense nuclear matter. Relativistic mean-field (RMF) models, suitable for describing both finite nuclei and high-density matter in NS, incorporate many-body interactions via mesons ($\sigma$, $\omega$, $\rho$). Two main RMF approaches describe nuclear properties: nonlinear meson terms in the Lagrangian density \cite{Boguta1977, Mueller1996, Steiner2004, Todd-Rutel2005} and density-dependent coupling parameters \cite{Typel1999, Typel2009, Lalazissis2005}. Other metamodels are constrained by low and high-density theoretical calculations \cite{Hebeler2013, Drischler2015, Kurkela2009}. Approaches to accommodate all possible EOSs include piecewise polytropic interpolation, speed of sound interpolation, spectral interpolation, and Taylor expansion \cite{Lindblom2012,Kurkela:2014vha,Most:2018hfd,LopeOter:2019pcq,Annala2019,Annala:2021gom}. 

Our study explores the Non-Linear model within the RMF framework. The $\sigma$-cut potential approach \cite{Maslov:2015lma}, proposing sharp increases in mean field self-interaction potential around the nuclear saturation density ($\rho_0$), effectively stiffens the EOS without affecting nuclear matter properties near $\rho_0$ \cite{Ma:2022fmu}. Few studies on the $\sigma$-cut scheme focus on EOS stiffness implications, including kaon condensation, hyperons in neutron stars \cite{Ma:2022fmu}, stellar properties, nuclear matter constraints \cite{PhysRevC.93.025806}, strangeness neutron stars within RMF \cite{Zhang:2018lpl}, NICER data analysis \cite{Kolomeitsev:2024gek}, and effects on pure nucleonic and hyperonic-rich NS matter \cite{Thakur:2024scc}. {The stiffening obtained with the introduction of the $\sigma$ cut potential may be seen as the inclusion of an exclusion volume effect \cite{Typel:2016srf}, or even as mimicking the onset of another baryonic phase such as the quarkyonic phase \cite{McLerran:2018hbz}.}

As upcoming observational data becomes increasingly refined, we are motivated to conduct a systematic study of the interior structure of neutron stars under three distinct scenarios. First, we consider neutron stars comprising only nucleonic degrees of freedom within the framework of the non-linear (NL) model. Second, we modify the NL model to include a $\sigma$-cut potential. Third, we investigate neutron stars that contain an admixture of dark matter, specifically focusing on fermionic dark matter for simplicity. Previous studies have indicated that the presence of dark matter tends to reduce both the mass and radius of neutron stars. Conversely, incorporating a $\sigma$-cut potential has been shown to increase these parameters. Given these contrasting effects, our goal is to calculate the Bayesian evidence for each of these three scenarios, thereby determining which model is most consistent with the latest observational data. By systematically evaluating these models, we aim to rank them based on their alignment with recent observations, providing a clearer understanding of the interior structure of neutron stars and the potential influence of dark matter and modified nucleonic interaction.

In the era of multimessenger astronomy, NS asteroseismology has become crucial for insights into dense matter EOS. Gravitational wave events like GW170817 \cite{LIGOScientific:2017vwq, LIGOScientific:2017ync} and GW190425, along with future detectors (LIGO-Virgo-KAGRA, Einstein Telescope \cite{Punturo:2010zz}, Cosmic Explorer), are expected to improve EOS determination. When an NS is perturbed, it oscillates in radial or non-radial modes. Radial modes involve simple expansion and contraction, maintaining a spherical shape. Non-radial oscillations (fundamental mode (f-mode), pressure mode (p-mode), and gravity mode (g-mode)) deviate from the spherical shape, driven by pressure and buoyancy \cite{Kokkotas:1999bd}. f-modes are significant for their gravitational radiation, detectable by current detectors. The 90\% credible interval for the f-mode frequency in GW170817 ranges from 1.43 kHz to 2.90 kHz for the more massive NS and from 1.48 kHz to 3.18 kHz for the less massive one \cite{Kunjipurayil:2022zah}. Recent studies have examined f-mode oscillations in NS using the Cowling approximation \cite{Pradhan:2020amo,Ranea-Sandoval:2018bgu,Dimmelmeier:2005zk}, which neglects background spacetime perturbations. These results were refined to include full general relativistic (GR) effects \cite{Pradhan:2022vdf,Kunjipurayil:2022zah}. Studies also explored f-modes in dark matter-admixed NS. Ref. \cite{Das:2021dru} examined the Higgs interaction mode of dark matter using the Cowling approximation, while Ref. \cite{Flores:2024hts} studied it in full GR. Ref. \cite{Shirke:2024ymc} investigated the neutron decay anomaly model of dark matter on f-mode oscillations within a full GR framework. This study focuses on non-radial oscillations of NS, such as $f$ and $p$ modes. Mode frequencies are calculated using the relativistic Cowling approximation \cite{Chirenti:2015dda,Kumar:2021hzo,Kumar:2023rut}. {This approximation is often used in the literature for a first-step calculation. The reason is the Cowling approximation can greatly simplify the pulsating equations in full GR simulations. The Cowling approximation might overestimate the frequency by 10-30\% depending on the compactness, in comparison to those frequencies obtained from linearized GR treatments \cite{10.1093/mnras/289.1.117}. {The overestimation in the case of Cowling compared to the linearized GR approach decreases with increasing stellar compactness. An explanation for this trend was suggested by \cite{Yoshida:1997wn}, indicating that as compactness increases, the influence of metric perturbations on the f-mode eigenfunction diminishes, as the eigenfunction is concentrated near the surface \cite{Chirenti:2015dda}.} }

We have performed a detailed statistical analysis using the current astrophysical observational data on NS properties within the Bayesian inference framework to explore the potential existence of dark matter in NS. Both baryonic (visible) matter and dark matter are considered within the RMF framework. {The EOSs are developed using empirical constraints based on experimental data regarding the properties of finite nuclei and observations from astrophysics. The nuclear matter properties taken into account include the pressure of symmetric nuclear matter ($P_{\rm SNM}$), the pressure resulting from symmetry energy ($P_{\rm sym}$), and the symmetry energy itself ($e_{\rm sym}$). These properties are empirically constrained across various densities using experimental data on the bulk characteristics of finite nuclei, such as nuclear masses, neutron skin thickness in $^{208}$Pb, dipole polarizability, isobaric analog states, and heavy ion collision (HIC) data covering the density range from $0.03$ to $0.32$ fm$^{-3}$. Additionally, astrophysical data utilized include the mass-radius posterior distributions for PSR J0030+0451\cite{Riley:2019yda,Miller:2019cac} and PSR J0740+6620\cite{Miller:2021qha, riley2021} as well as the posterior distribution for dimensionless tidal deformability for components of binary neutron stars from the GW170817 event.}  We examined three different scenarios: i) solely nucleonic degrees of freedom within RMF, referred to as NL, ii) an EOS stiffened by a modified $\sigma$-cut potential, referred to as NL-$\sigma$ cut, and iii) {impact of dark matter on neutron stars, specifically, how the presence of fermionic dark matter in admixed NS matter through neutron decay {affects the nuclear} equation of state, referred as NL-DM.} 
We investigated which of these three scenarios aligns better with recent observational data on NS. Additionally, we analyzed the impact of PREX-II experimental data with or without inclusion in all cases, which often contradicts various other nuclear data reported in the literature. This study provides a definitive answer to the probabilistic measure of the possibility of dark matter existence in NS.
 
The structure of this paper is as follows: Section \ref{formalishm} covers the theoretical framework used in this study. The results are discussed in Section \ref{results}. Lastly, the conclusions and main insights of the study are presented in Section \ref{conclussion}.

\section{Methodology \label{formalishm}}
The development of the EOS will be carried out within the Relativistic Mean Field (RMF) framework. Our analysis will focus on the RMF model that includes non-linear mesonic interactions (labeled as NL), modifications to the $\sigma$ potential (referred to as NL $\sigma$ cut), and the impacts of dark matter in the NL framework (NL DM). We will employ Bayesian inference, using the latest constraints from nuclear and astrophysical observations. Utilizing the resulting posterior distribution, we will examine the non-radial oscillations, such as the fundamental mode (f-mode) and the pressure mode (p-mode), across all three variations of the model. The details of the framework are as follows. 

\subsection{NL model}
In the RMF model, the EOS for nuclear matter is described by the interaction of the scalar-isoscalar meson $\sigma$, the vector-isoscalar meson $\omega$, and the vector-isovector meson $\varrho$. The Lagrangian density is given by \citep{Fattoyev:2010mx,Dutra:2014qga,Malik:2023mnx}
        \begin{equation}
          \mathcal{L}=   \mathcal{L}_N+ \mathcal{L}_M + \mathcal{L}_{NL} +\mathcal{L}_{leptons}
        \end{equation} 
with
$$\mathcal{L}_{N} = \bar{\Psi}\Big[\gamma^{\mu}\left(i \partial_{\mu}-g_{\omega} \omega_\mu - \frac{1}{2}g_{\varrho} {\boldsymbol{\tau}} \cdot \boldsymbol{\varrho}_{\mu}\right) - \left(m_N - g_{\sigma} \sigma\right)\Big] \Psi$$
represents the Dirac Lagrangian density for the neutron and proton doublet with a bare mass $m_N$, where $\Psi$ denotes a Dirac spinor, $\gamma^\mu$ are the Dirac matrices, and $\boldsymbol{\tau}$ symbolizes the isospin operator.
The $\mathcal{L}_{M}$ is the Lagrangian density for the mesons, given by
\begin{eqnarray}
\mathcal{L}_{M}  &=& \frac{1}{2}\left[\partial_{\mu} \sigma \partial^{\mu} \sigma-m_{\sigma}^{2} \sigma^{2} \right] - \frac{1}{4} F_{\mu \nu}^{(\omega)} F^{(\omega) \mu \nu} + \frac{1}{2}m_{\omega}^{2} \omega_{\mu} \omega^{\mu}   \nonumber \\
  &-& \frac{1}{4} \boldsymbol{F}_{\mu \nu}^{(\varrho)} \cdot \boldsymbol{F}^{(\varrho) \mu \nu} + \frac{1}{2} m_{\varrho}^{2} \boldsymbol{\varrho}_{\mu} \cdot \boldsymbol{\varrho}^{\mu} \nonumber
\end{eqnarray}
where $F^{(\omega, \varrho)\mu \nu} = \partial^ \mu A^{(\omega, \varrho)\nu} -\partial^ \nu A^{(\omega, \varrho) \mu}$ are the vector meson  tensors, and
\begin{eqnarray}
\mathcal{L}_{NL}&=&-\frac{1}{3} b~m_N~ g_\sigma^3 (\sigma)^{3}-\frac{1}{4} c (g_\sigma \sigma)^{4}+\frac{\xi}{4!} g_{\omega}^4 (\omega_{\mu}\omega^{\mu})^{2}  \nonumber \\
&+&\Lambda_{\omega}g_{\varrho}^{2}\boldsymbol{\varrho}_{\mu} \cdot \boldsymbol{\varrho}^{\mu} g_{\omega}^{2}\omega_{\mu}\omega^{\mu} \nonumber
\end{eqnarray}
includes the non-linear mesonic terms characterized by the parameters $b$, $c$, $\xi$, and $\Lambda_{\omega}$, which manage the high-density properties of nuclear matter. The coefficients $g_i$ represent the couplings between the nucleons and the meson fields $i = \sigma, \omega, \varrho$, which have masses denoted by $m_i$.

Finally, the dynamics of leptons are described by the Lagrangian density  
$$\mathcal{L}_{leptons}= \bar{\Psi_l}\Big[\gamma^{\mu}\left(i \partial_{\mu}  
-m_l \right)\Big]\Psi_l,$$ 
where $\Psi_l~(l= e^-, \mu^-)$ denotes the lepton spinor for electrons and muons; leptons are considered non-interacting.

The energy density of the baryons and  leptons is given by the following expressions:
\begin{equation}
\begin{aligned}
\epsilon &= \sum_{i=n,p,e,\mu}\frac{1}{\pi^2}\int_0^{k_{Fi}} \sqrt{k^2+{m_i^*}^2}\, k^2\, dk \\
&+ \frac{1}{2}m_{\sigma}^{2}{\sigma}^{2}+\frac{1}{2}m_{\omega}^{2}{\omega}^{2}+\frac{1}{2}m_{\varrho}^{2}{\varrho}^{2}\\
&+ \frac{b}{3} m_{\rm N} (g_{\sigma}{\sigma})^{3}+\frac{c}{4}(g_{\sigma}{\sigma})^{4}+\frac{\xi}{8}(g_{\omega}{\omega})^{4} + \Lambda_{\omega}(g_{\varrho}g_{\omega}{\varrho}{\omega})^{2},
\end{aligned}
\end{equation}
where $m_i^*=m_i-g_{\sigma} \sigma$ for protons and neutrons, $m_i^*=m_i$ for electrons and muons, and $k_{Fi}$ is the Fermi moment of particle $i$.
{The $\sigma$, $\omega$ and $\rho$ are the  mean-filed values of the corresponding mesons \cite{Malik:2023mnx}}.

Once we have the energy density for a given EOS model, we can compute the chemical potential of neutron ($\mu_n$) and proton ($\mu_p$). The chemical potential of electron ($\mu_e$) and muon ($\mu_\mu$) can be computed using the condition of $\beta$-equilibrium : $\mu_n-\mu_p=\mu_e$ and $\mu_e$ = $\mu_\mu$ and the charge neutrality: $\rho_p =\rho_e + \rho_\mu$. Where $\rho_e$ and $\rho_\mu$ are the electron and muon number density. Furthermore, using the thermodynamic relation, we can obtain the pressure, 
\begin{equation}
P = \sum_{i}\mu_{i}\rho_{i}-\epsilon.
\label{eq_9}
\end{equation}

\subsection{NL $\sigma$ Cut}
We further investigate the addition of the $\sigma$ cut potential $U_{cut}(\sigma)$ to the RMF Lagrangian, as mentioned in \cite{Maslov:2015lma,Zhang:2018lpl,Patra:2022lds}. The $U_{cut}(\sigma)$ has a logarithmic form, as in \cite{Maslov:2015lma}, which only affects the $\sigma$ field at high density and is given by, 
\begin{eqnarray}
U_{cut}(\sigma) = \alpha \ln [ 1 + \exp\{\beta(g_{\sigma}\sigma/m_{N}-f_{s})\}]
\end{eqnarray}
where $\alpha$ = $m_{\pi}^{4}$ and $\beta$ = 120 \cite{Maslov:2015lma}, and the parameter $f_{s}$ is determined by Bayesian inference.

The Lagrangian density equation (1) for $\mathcal{L}_{NL,\sigma \rm cut}$  with $U_{cut}(\sigma)$ is given by,
\begin{eqnarray}
    \mathcal{L}_{NL,\sigma cut}&=&\mathcal{L}_{NL} - U_{cut}(\sigma)
\end{eqnarray}
and the  meson fields are determined from the  equations 
		\begin{eqnarray}
			{\sigma}&=& \frac{g_{\sigma}}{m_{\sigma,{\rm eff}}^{2}}\sum_{i} \rho^s_i\label{sigma}\\
			{\omega} &=&\frac{g_{\omega}}{m_{\omega,{\rm eff}}^{2}} \sum_{i} \rho_i \label{omega}\\
			{\varrho} &=&\frac{g_{\varrho}}{m_{\varrho,{\rm eff}}^{2}}\sum_{i} I_{3} \rho_i, \label{rho}
		\end{eqnarray}
 where $\rho^s_i$ and $\rho_i$ are, respectively, the scalar density and the number density of nucleon $i$, and
 \begin{eqnarray}
     m_{\sigma,{\rm eff}}^{2}&=& m_{\sigma}^{2}+{ b ~m_N ~g_\sigma^3}{\sigma}+{c g_\sigma^4}{\sigma}^{2} + \frac{U_{cut}^{'}(\sigma)}{\sigma}\\
    m_{\omega,{\rm eff}}^{2}&=& m_{\omega}^{2}+ \frac{\xi}{3!}g_{\omega}^{4}{\omega}^{2} +2\Lambda_{\omega}g_{\varrho}^{2}g_{\omega}^{2}{\varrho}^{2}\label{mw}\\
    m_{\varrho,{\rm eff}}^{2}&=&m_{\varrho}^{2}+2\Lambda_{\omega}g_{\omega}^{2}g_{\varrho}^{2}{\omega}^{2} \label{mr}
 \end{eqnarray}
where $U_{cut}^{'}(\sigma)$ is the derivative of  $U_{cut}(\sigma)$ with respect to $\sigma$.
In these equations, the meson fields should be interpreted as their expectation values. The energy density and pressure for this case are \cite{Maslov:2015lma}
\begin{equation}
\begin{aligned}
\epsilon_{\sigma \rm cut} &=  \epsilon +U_{cut}(\sigma)
\end{aligned}
\end{equation}

\begin{equation}
\begin{aligned}
P_{\sigma \rm cut} &=  P - U_{cut}(\sigma).
\end{aligned}
\end{equation}

\subsection{Dark Matter}
The current interpretation of experimental data on neutron decay suggests the potential presence of phenomena beyond the standard model of physics~\cite{PhysRevLett.120.202002,PhysRevLett.120.191801,PhysRevLett.121.061801}. Neutrons predominantly undergo $\beta$-decay:
$$n \rightarrow p + e^{-} + \bar{\nu_{e}}~.$$ The two experiments that measure neutron lifetime, namely the beam experiment and the bottle experiment, yield two different neutron lifetimes. The bottle experiment yields $\tau_{\rm bottle}=879.6\pm0.6$ s~\cite{Mampe:1993an,Serebrov:2004zf,Pichlmaier:2010zz,Steyerl:2012zz,Arzumanov:2015tea,Pattie:2017vsj}, and the beam experiment measurement gives $\tau_{\rm beam}=888.0\pm 2.0$ s~\cite{Yue:2013qrc,Byrne:1996zz}. These two neutron lifetime measurements differ by $4\sigma$, indicating a need to reconcile our understanding of fundamental interactions. In Ref.~\cite{PhysRevLett.120.191801}, authors have come up with an intriguing suggestion where they propose that new decay channels of neutrons into dark matter particles could account for the anomaly in the neutron lifetime measurement. These new decay channels, where neutrons decay into dark matter particles, could be potentially interesting for neutron star physics. Several recent studies have investigated this possibility, suggesting that neutron stars can serve as powerful laboratories to test the proposed decay of neutrons into dark matter particles\cite{Husain:2022bxl,Bastero-Gil:2024kjo,PhysRevLett.121.061801,Shirke:2023ktu,Motta:2018bil,Shirke:2024ymc}.
In this work, we examine the effect of neutron decay on neutron star dynamics using the decay channel involving baryon-number-
violating beyond the standard model (BSM) interaction,
\begin{equation}\label{eqn:darkdecay}
n\rightarrow\chi+\phi~,
\end{equation}
where $\chi$ is a dark spin-1/2 fermion, and $\phi$ is a light dark boson. Other decay channels of neutrons are also possible, e.g., $n\rightarrow \chi+\gamma$, and $n\rightarrow \chi+e^+e^-$. However, phenomenologically all decay channels are not favored; e.g., laboratory experiment puts stringent constraints on the decay channel $n\rightarrow \chi+\gamma$~\cite{Tang:2018eln}. The decay channel $n \rightarrow \chi + \phi$ is especially intriguing in the context of neutron star physics, as it can be argued that the light dark matter boson $\phi$ would quickly escape the neutron star, rendering it insignificant. Conversely, some of the neutrons within the neutron star will transform into fermionic dark matter $\chi$ due to the BSM interaction. Physically these dark matter particles will experience the gravitational potential of the neutron star and will reach thermal equilibrium with the surrounding neutron star matter. This sets the equilibrium condition 
\begin{equation}\label{eqn:chemicalequilibrium}
    \mu_{\chi}=\mu_{n}~.
\end{equation}

Nuclear stability requires \( 937.993 \, \text{MeV} < m_{\chi} + m_{\phi} < m_n = 939.565 \, \text{MeV} \) \cite{Motta:2018bil,Shirke:2023ktu}. For the dark particles to remain stable and avoid further beta decay, the condition \( |m_{\chi} - m_{\phi}| < m_p + m_e = 938.783 \, \text{MeV} \) must be met \cite{Fornal:2020gto}.

To account for dark matter (DM) self-interactions, we introduce vector interactions between dark particles, described by:
\begin{equation}
    \mathcal{L} \supset -g_V \Bar{\chi} \gamma^{\mu} \chi V_{\mu} - \frac{1}{4} V_{\mu\nu} V^{\mu\nu} + \frac{1}{2} m_V^2 V_{\mu} V^{\mu}~,
\end{equation}
where \( g_V \) is the coupling strength and \( m_V \) is the mass of the vector boson. This introduces an additional interaction term in the energy density, beyond the free fermion part. The energy density of DM is given by:
\begin{equation}
    \epsilon_{DM} = \frac{1}{\pi^2} \int_{0}^{k_{F_{\chi}}} k^2 \sqrt{k^2 + m_{\chi}^2} \, dk + \frac{1}{2} G_{\chi} n_{\chi}^2,
    \label{eqn:endens_dm}
\end{equation}
where,
\begin{equation} \label{eqn:Gdefinition}
    G_{\chi} = \left( \frac{g_V}{m_V} \right)^2, \qquad n_{\chi} = \frac{k_{F_{\chi}}^3}{3 \pi^2}
\end{equation}
and
\[
\mu_\chi = \sqrt{k^2 + m_\chi^2} + G_{\chi} n_\chi.
\]
\subsection{\textbf{Bayesian Likelihood}}
The Bayesian likelihood is fundamental in Bayesian statistics \cite{Imam:2024gfh}, enabling the probability assessment of a hypothesis to be revised in light of new data or evidence. Within Bayesian inference, the posterior distribution illustrates the credibility of parameter values based on the observed data.

{\it Data}:- Within our inference analysis, we have taken into account various constraints ranging from nuclear physics experiments to astrophysical observations. The data set is presented in Table \ref{tab1}. {The constraints are established based on empirical data derived from experimental observations of finite nuclei properties, such as nuclear masses, neutron skin thickness in $^{208}$Pb, dipole polarizability, and isobaric analog states, alongside heavy ion collision (HIC) data spanning densities from $0.03$ to $0.32$ fm$^{-3}$.  Moreover, our analysis includes astrophysical data like the mass-radius relationships observed in pulsars PSR J0030+0451 and PSR J0740+6620, as well as the tidal deformability of binary neutron star systems as demonstrated by the GW170817 event.} We apply these constraints within a Bayesian framework to refine the EOSs. The terrestrial and astrophysical data sources used in our analysis are listed in Table \ref{tab1}.

{\it Likelihood}:- The Likelihood of the given data is described as follows:
\begin{itemize}
    \item \textbf{Experimental data:} For experimental data, \( D_{\text{expt}} \pm \sigma \) having a symmetric Gaussian distribution, the likelihood is given as,

\[
L(D_{\text{expt}}|\theta) = \frac{1}{\sqrt{2\pi\sigma^2}} \exp\left(-\frac{(D(\theta) - D_{\text{expt}})^2}{2\sigma^2}\right) = L_{\text{expt}}.
\]

Here, \( D(\theta) \) is the model value for a given model parameter set \(\theta\).
  \item \textbf{GW observation}: For GW observations, information about EOS parameters comes from the masses $m_1, m_2$ of the two binary components and the corresponding tidal deformabilities $\Lambda_1, \Lambda_2$. In this case, 
 \begin{align}
    P(d_{\mathrm{GW}}|\mathrm{EOS}) = \int^{M_u}_{m_l}dm_1 \int^{m_1}_{M_l} dm_2 P(m_1,m_2|\mathrm{EOS})   \nonumber \\
    \times P(d_{\mathrm{GW}} | m_1, m_2, \Lambda_1 (m_1,\mathrm{EOS}), \Lambda_2 (m_2,\mathrm{EOS})) \nonumber \\
    =\mathcal{L}^{\rm GW}
    \label{eq:GW-evidence}
 \end{align}
where P(m$|$EOS) ~\citep{Agathos_2015,Wysocki-2020,Landry_2020PhRvD.101l3007L,Biswas:2020puz} can be written as, 
\begin{equation}
    P(m|\rm{EOS}) = \left\{ \begin{matrix} \frac{1}{M_u - M_l} & \text{ iff } & M_l \leq m \leq M_u, \\ 0 & \text{ else, } & \end{matrix} \right.
\end{equation}
In our calculation, we set $M_l$ = 1 M$_{\odot}$ and $M_u$ as the maximum mass for a given EOS. 

\item \textbf{X-ray observation(NICER)}: X-ray observations give the mass and radius measurements of NS. Therefore, the corresponding evidence takes the following form,
\begin{align}
    P(d_{\rm X-ray}|\mathrm{EOS}) = \int^{M_u}_{M_l} dm P(m|\mathrm{EOS}) \nonumber \\ \times
    P(d_{\rm X-ray} | m, R (m, \mathrm{EOS})) \nonumber \\
    = \mathcal{L}^{\rm NICER}.
\end{align}
 \\
Here again, M$_l$ represents a mass of 1 M$_\odot$, and M$_u$ denotes the maximum mass of a neutron star according to the respective EOS.\\

\vspace{0.5cm}
The final likelihood for the three scenarios:\\
\begin{equation}
    \mathcal{L} = 
    \mathcal{L}^{\rm EXPT}
    \mathcal{L}^{\rm GW}
    \mathcal{L}^{\rm NICER I}
    \mathcal{L}^{\rm NICER II}\,.
    \label{eq:finllhd}
\end{equation}
NICER I and NICER II refer to the mass-radius measurements of the pulsars PSR J0030+0451 and PSR J0740+6620, respectively.
\end{itemize}

{We have successfully implemented the nested sampling algorithm using \texttt{PyMultiNest} \cite{Buchner:2014nha}, setting the number of live points at 2000. This configuration allowed us to obtain a robust posterior distribution with approximately 9,000 samples, derived from roughly 400,000 likelihood evaluations. Each individual case costs about 10,000 CPU hours on a high-performance computing system, \texttt{DEUCALION} \cite{deucalion2024}.}

\subsection{Non-radial oscillation modes}
Our calculation of non-radial modes oscillations is restricted to the well-celebrated Cowling approximation, which neglects the perturbations in the background metric. 

In the Cowling approximation, the spacetime metric for a spherically symmetric background is given by
\begin{equation}
     ds^2=-e^{2\Phi (r)}dt^2+e^{2\Lambda (r)}dr^2+r^2 d\theta^2+r^2\sin^2{\theta} d\phi^2. \label{eqn:metric}
 \end{equation}
 
In order to find mode frequencies, one has to solve the following differential equations (\cite{Pradhan:2020amo}):
\\
\begin{eqnarray}
    \frac{d W(r)}{dr}&=&\frac{d \epsilon}{dp}\left[\omega^2r^2e^{\Lambda (r)-2\phi (r)}V (r)+\frac{d \Phi(r)}{dr} W (r)\right] \nonumber \\
    &-& l(l+1)e^{\Lambda (r)}V (r) \nonumber \\
    \frac{d V(r)}{dr} &=& 2\frac{d\Phi (r)}{dr} V (r)-\frac{1}{r^2}e^{\Lambda (r)}W (r) 
    \label{eqn:perteq}
\end{eqnarray}
where,
\begin{equation*}
    \frac{d \Phi(r)}{dr}=\frac{-1}{\epsilon(r)+p(r)}\frac{dp}{dr}.
\end{equation*}
The solution of Eq.~(\ref{eqn:perteq}) with the fixed background metric Eq.~(\ref{eqn:metric}) near the origin will behave as follows:
\begin{equation}
    W (r)=Ar^{l+1}, \> V (r)=-\frac{A}{l} r^l.
\end{equation}
The vanishing perturbed Lagrangian pressure at the surface will provide another constraint to be included while solving Eq.~\ref{eqn:perteq}, which is given by,
\begin{equation} 
    \omega^2e^{\Lambda (R)-2\Phi (R)}V (R)+\frac{1}{R^2}\frac{d\Phi (r)}{dr}\Big|_{r=R}W (R)=0.
\label{eqn:bc}
\end{equation}
Eqs.~(\ref{eqn:perteq}) are eigenvalue equations. Among the solutions, those that satisfy the boundary condition given by Eq.~(\ref{eqn:bc}) are the eigenfrequencies of the star.

\begin{table*}
\caption{\label{tab1}The empirical values of symmetry energy (e$_{\rm sym}$), symmetry energy pressure (P$_{\rm sym}$), and symmetric nuclear matter pressure (P$_{\rm SNM}$) from experimental data on the bulk properties of finite nuclei and HIC. The astrophysical observational constraints on the radii and tidal deformability of neutron stars. See Ref\cite{Tsang2024} for details.}
\label{tab:constraints}
\begin{tabular*}{\textwidth}{@{\extracolsep{\fill}}lcccc}
\hline
Symmetric matter  &&&& \\
Constraints & $n$ (fm$^{-3}$) & P$_\text{SNM}$ (MeV/fm$^3$)&  & Ref. \\
\hline
%\hline
HIC(DLL)  & 0.32             & $10.1\pm3.0$                 &                      &\cite{Danielewicz:2002pu} \\
HIC(FOPI)  & 0.32                & $10.3\pm2.8$                 &                      &\cite{LeFevre:2015paj} \\ \\
\hline
\hline
Asymmetric matter  &&&& \\
Constraints & $n$ (fm$^{-3}$) & S($n$) (MeV)      & P$_\text{\rm sym}$ (MeV/fm$^3$) & Ref. \\ 
\hline
Nuclear structure &&&&\\
$\alpha_D$    & 0.05           & $15.9\pm1.0$ &                                        &~\cite{Zhang:2015ava}        \\
PREX-II       & 0.11             &              & $2.38\pm0.75$           &~\cite{PREX:2021umo,Reed:2021nqk,decoding2022}\\
\\
Nuclear masses &&&&\\ 
Mass(Skyrme)  & 0.101          & $24.7\pm0.8$ &                                          &~\cite{brown2013constraints,decoding2022}
\\
Mass(DFT) &  0.115             & $25.4\pm1.1$ &                                             &~\cite{kortelainen2012nuclear,decoding2022}  \\
IAS           & 0.106           & $25.5\pm1.1$ &                                             &~\cite{danielewicz2017symmetry,decoding2022}     \\
\\
Heavy-ion collisions \\ 
HIC(Isodiff)  & 0.035       & $10.3\pm1.0$ &                                             &~\cite{tsang2009constraints,decoding2022}    \\
HIC(n/p ratio) & 0.069\          & $16.8\pm1.2$ &                                         &~\cite{morfouace2019constraining,decoding2022}\\
HIC($\pi$)    & 0.232           & $52\pm13$    & $10.9\pm8.7$                         &~\cite{estee2021probing,decoding2022}    \\

HIC(n/p flow) & 0.240           &              & $12.1\pm8.4$          &~\cite{cozma2018feasibility,RUSSOTTO2011471,Russotto16,decoding2022} \\
\\
\hline
\hline
\end{tabular*}
\begin{tabular*}{\textwidth}{@{\extracolsep{\fill}}lcccc}
Astrophysical    &&&& \\
Constraints & $M_\odot$ & R (km) & $\Lambda_{1.36}$ & Ref.\\
\hline 
LIGO~\footnote{LVK collaboration,~\href{https://dcc.ligo.org/LIGO-P1800115/public}{https://dcc.ligo.org/LIGO-P1800115/public}} & 1.36 & & $300_{-230}^{+420}$ &~\cite{Abbott18a} \\ \vspace{0.2mm}
*Riley ~PSR J0030+0451~\footnote{~\href{https://zenodo.org/records/8239000}{https://zenodo.org/records/8239000}}& 1.34 & $12.71^{+1.14}_{-1.19}$ & &~\cite{Riley:2019yda} \\\vspace{0.2mm}

*Miller PSR J0030+0451~\footnote{~\href{https://zenodo.org/record/3473466\#.XrOt1nWlxBc}{https://zenodo.org/record/3473466\#.XrOt1nWlxBc}}& 1.44 & $13.02^{+1.24}_{-1.06}$ & &~\cite{Miller:2019cac} \\\vspace{0.2mm}
%\hline
*Riley ~PSR J0740+6620~\footnote{~\href{https://zenodo.org/records/4697625}{https://zenodo.org/records/4697625}} & 2.07 & 12.39$^{+1.30}_{-0.98}$ & &~\cite{riley2021}\\\vspace{0.2mm}
*Miller PSR J0740+6620~\footnote{~\href{https://zenodo.org/records/4670689}{https://zenodo.org/records/4670689}} & 2.08 & $13.7~^{+2.6}_{-1.5}$ & &~\cite{Miller:2021qha}\\\vspace{0.2mm}
%\hline
*Choudhury PSR J0437-4715~\footnote{~\href{https://zenodo.org/records/12703175}{https://zenodo.org/records/12703175}} & 1.418 & $11.36~^{+0.95}_{-0.63}$ & &~\cite{Choudhury:2024xbk}

\\
\hline
\end{tabular*}
\end{table*}

\section{Results \label{results}}
{We analyze various neutron star properties considering three different scenarios: i) neutron stars (NS) with only nucleonic degrees of freedom, modeled using the Relativistic Mean Field (RMF) approach with mesonic nonlinear interaction (NL), ii) NS with nucleonic degrees of freedom in the RMF model but with a modified $\sigma$ potential (NL-$\sigma$ cut), and iii) NS with admixed dark matter, where the nucleonic matter is modeled with NL and the dark matter is produced by the neutron decay channel as mentioned in Ref. \cite{PhysRevLett.120.191801} (NL-DM).  The detailed formalism for all three scenarios can be found in the previous section. The parameter spaces of the model for all these cases are explored using the Bayesian inference framework. Various constraints from nuclear physics experiments that involve both symmetric and asymmetric matter are considered, as well as astrophysical constraints on the properties of NS, such as NICER radius measurements and GW170817 tidal deformability, all summarized in Table \ref{tab:constraints} including references (for more details see ref. \cite{Tsang2024}). Each scenario is also examined with additional constraints from PREX II \cite{PREX:2021umo,Reed:2021nqk}. Finally, we examined the nonradial oscillations of neutron stars (NS), focusing on $f$ and $p$ mode oscillations and using the posterior obtained for each case.} 

\begin{figure*}
    \centering
    \includegraphics[width=0.47\textwidth]{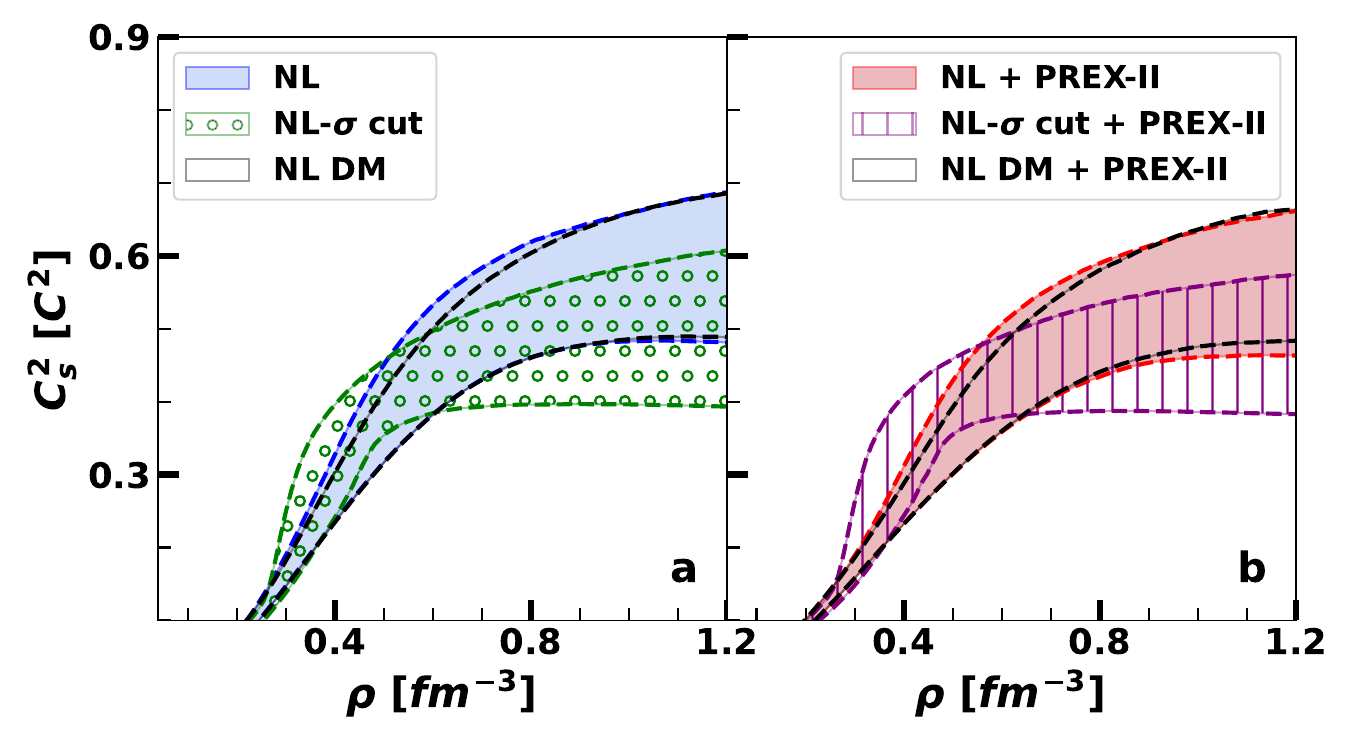}
    \includegraphics[width=0.47\textwidth]{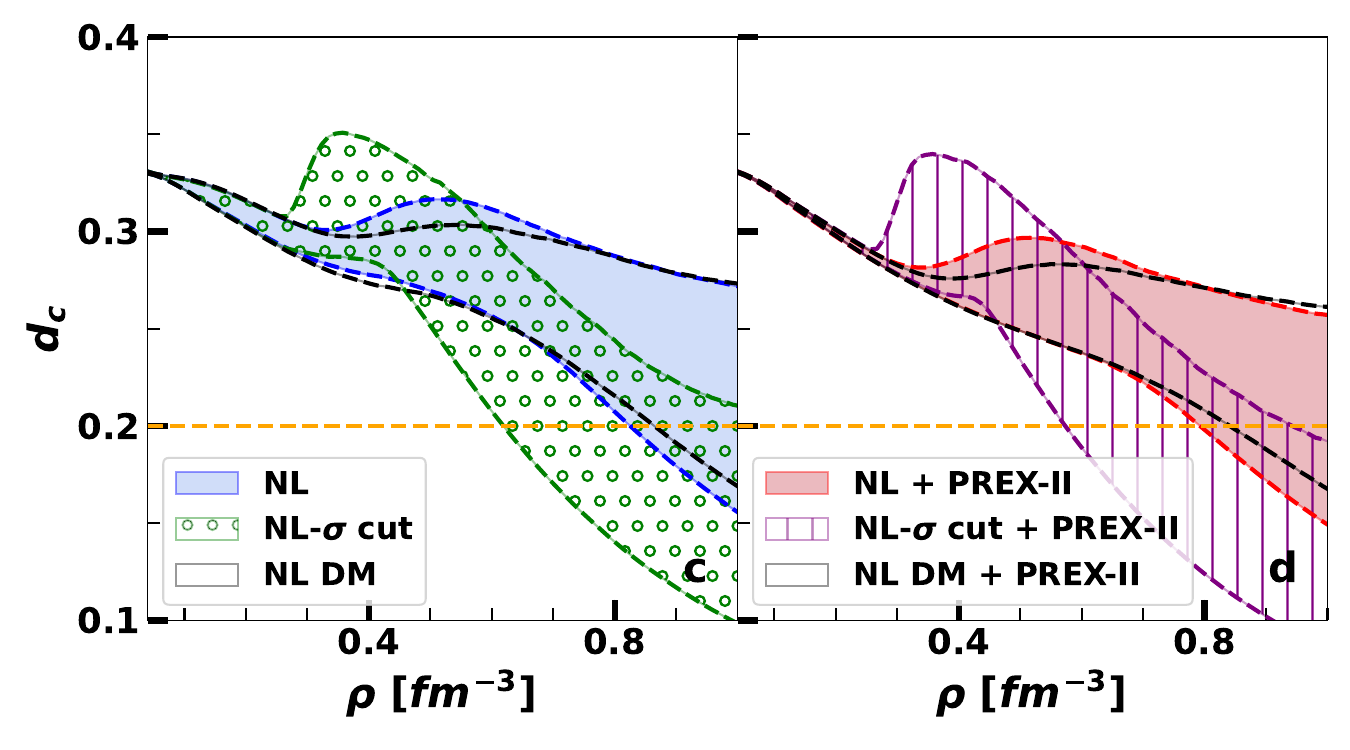}
    \caption{(left) 90\% confidence intervals for $c_s^2$ vs. baryon density for {NL, NL-$\sigma$ cut and NL-DM} models. (right) 90\% confidence intervals for $d_c$ vs. baryon density for {NL, NL-$\sigma$ cut and NL-DM} models. $d_c = \sqrt{\Delta^2 + \Delta'^2}$, where $\Delta' = c_s^2 \left( \frac{1}{\gamma} - 1 \right)$, and $\Delta = \frac{1}{3} - \frac{P}{\epsilon}$. The left (right) panel subfigure "a" excludes PREX-II data; the subfigure "b" includes it.}
    \label{fig:vs-density}
\end{figure*}

\begin{figure*}[t]
    \centering
    \includegraphics[width=0.48\textwidth]{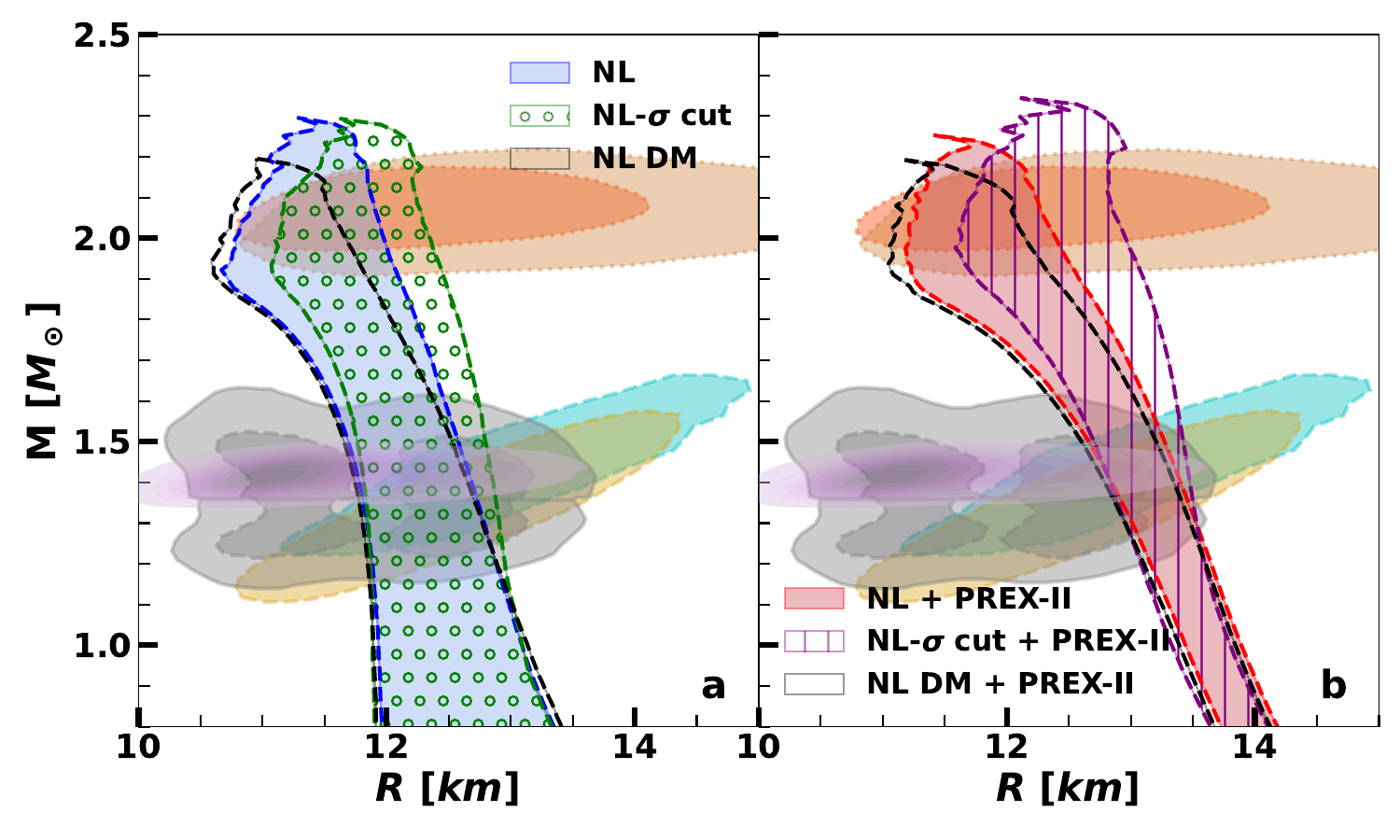}
    \includegraphics[width=0.48\textwidth]{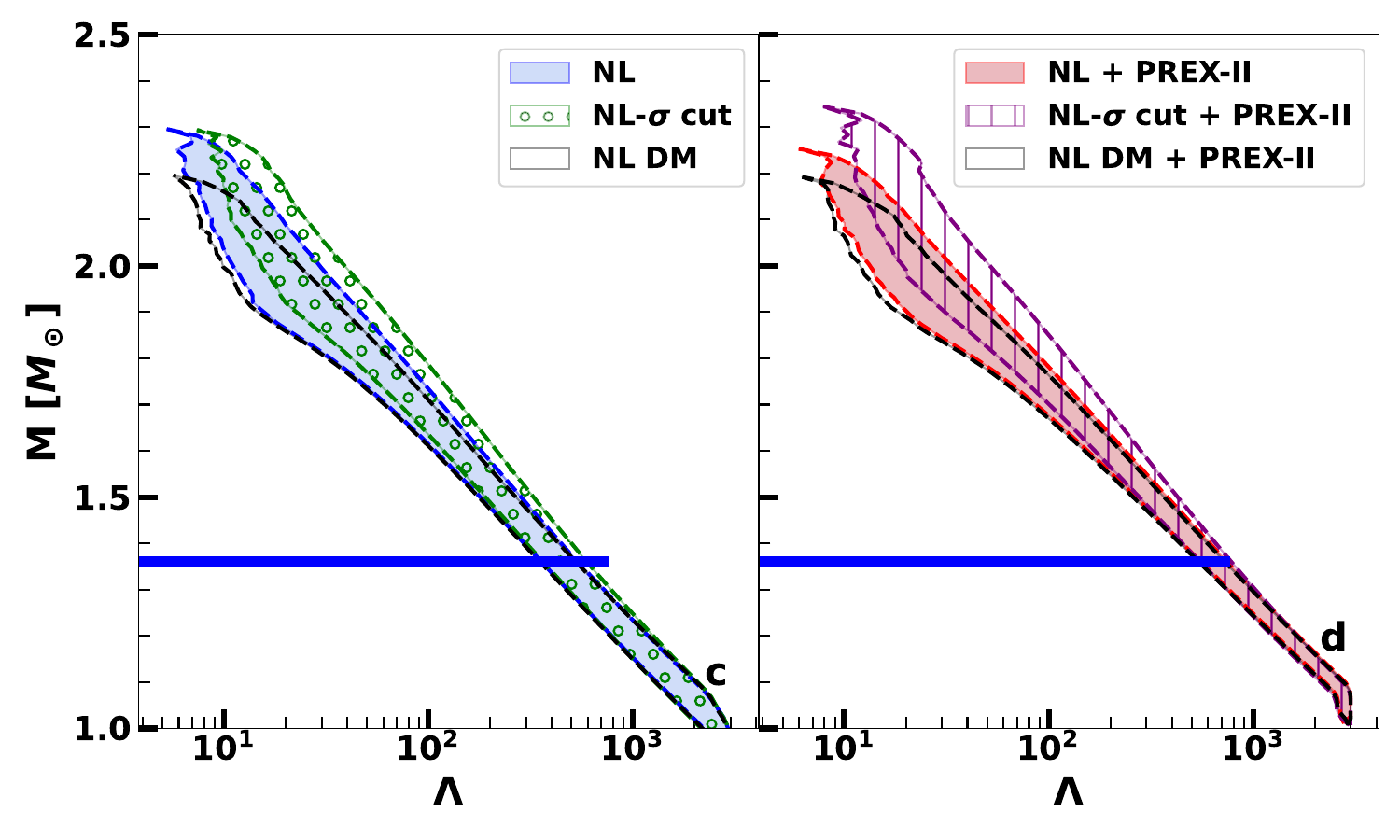}
    \caption{(left) The 90\% credible interval (CI) region for the neutron star (NS) mass-radius posterior $P(R|M)$ is plotted for the NL, NL-$\sigma$ cut, and NL-DM models. The gray area indicates the constraints obtained from the binary components of GW170817, with their respective 90\% and 50\% credible intervals. Additionally, the plot includes the 1 $\sigma$ (68\%) CI for the 2D mass-radius posterior distributions of the millisecond pulsars PSR J0030 + 0451 (in cyan and yellow color) \cite{Riley:2019yda, Miller:2019cac} and PSR J0740 + 6620 (in orange and peru color)\cite{Riley:2021pdl, Miller:2021qha}, based on NICER X-ray observations. Furthermore, we display the latest NICER measurements for the mass and radius of PSR J0437-4715 \cite{Choudhury:2024xbk} (lilac color). (right) The 90\% CI region for the mass-tidal deformability posterior $P(\Lambda|M)$ for the same models is presented. The blue bars represent the tidal deformability constraints at 1.36 $M_{\odot}$ \cite{LIGOScientific:2018cki}. In the left (right) subfigure, panels a(c) and b(d) correspond to the data without and with PREX-II data, respectively.}
    \label{fig:mass-radius-lam-DM}
\end{figure*}

{Fig. \ref{fig:vs-density} depicts the 90\% confidence intervals (CI) for the squared speed of sound ($c_s^2$) versus baryon density ($\rho$) for different models, shown in the left panel (subfigures a and b). Subfigure $a$ excludes PREX-II data, while subfigure $b$ includes it. The PREX-II data have an insignificant effect on $c_s^2$ since this data primarily affects the density-dependent symmetry energy, and the squared speed of sound is largely affected by the symmetric nuclear matter part of the EOS. Among the scenarios: NL, NL-$\sigma$ cut, and NL-DM, the presence of dark matter in the NL-DM model slightly affects the NL model at intermediate densities (0.4-0.8 fm$^{-3}$), reducing $c_s^2$ and thus softening the EOS. For the NL-$\sigma$ cut model, modifications to the $\sigma$ potential result in a stiffer $c_s^2$ at low densities below 0.5 fm$^{-3}$, becoming softer at higher densities compared to both the NL and the {NL-DM} models. This occurs because, in the NL-$\sigma$ cut model, the $\sigma$ {field saturates due to a very stiff $\sigma$ potential, and the effective mass stays at a constant value that can be above 0.5~$m_{N}$. The repulsive $\omega$ potential becomes dominant, and at high densities, the effect of the $\omega^4$ term softens the EOS.}
}

\begin{figure*}
    \centering
    \includegraphics[width=0.45\textwidth]{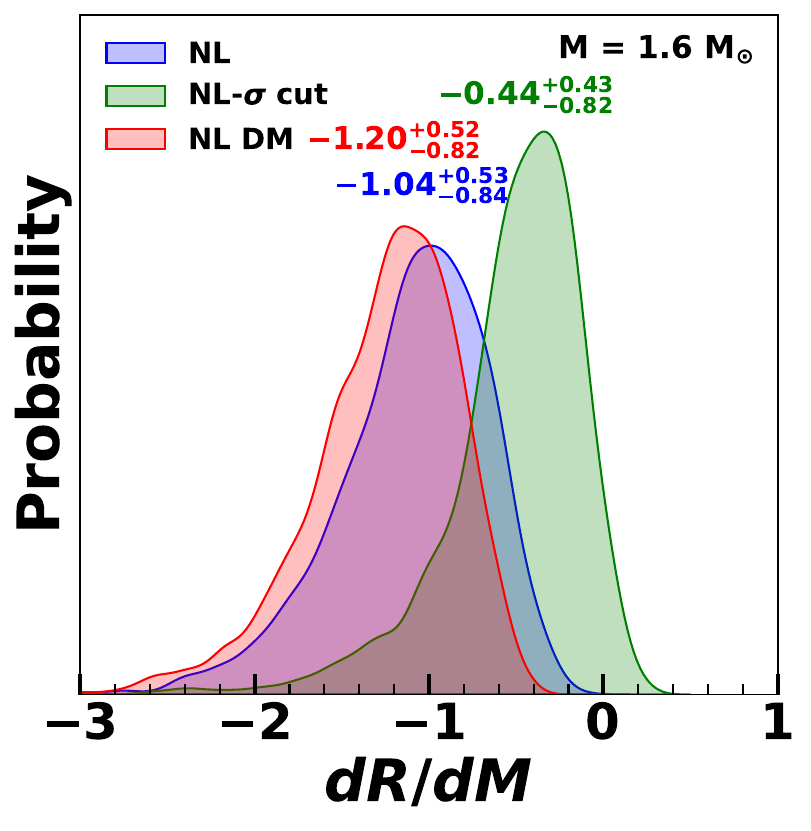}
    \includegraphics[width=0.45\textwidth]{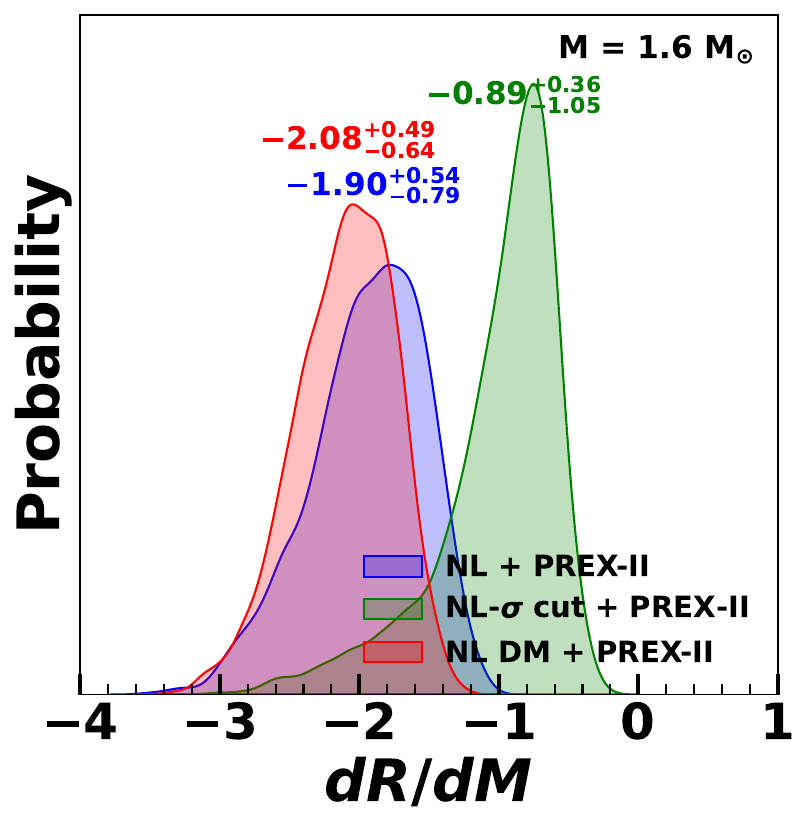}
    \caption{ The $dM/dR$ distribution at a neutron star mass of 1.6 $M_\odot$ for three scenarios: NL, NL-$\sigma$ cut, and NL-DM, shown without PREX-II data on the left and with PREX-II data on the right.}
    \label{fig:dm_dr_noprex}
\end{figure*}

{The parameter $d_c$, introduced in Ref. \cite{Annala:2023cwx}, is defined as $d_c = \sqrt{\Delta^2 + (\Delta')^2}$,
{$\Delta=\frac{1}{3}-\frac{P}{\epsilon}$ is the renormalized trace anomaly introduced in \cite{Fujimoto:2022ohj},}
 $\Delta' = c_s^2 \left(\frac{1}{\gamma} - 1\right)$ denotes the logarithmic derivative of $\Delta$, and $\gamma = \frac{d \ln P}{d \ln \epsilon}$. In the conformal limit, $c_s^2$ and $\gamma$ reach $1/3$ and $1$ respectively. It has been proposed that a value $d_c \lesssim 0.2$ suggests proximity to the conformal limit, as both $\Delta$ and its derivative need to be small for this to hold true. Since quark matter is expected to show approximate conformal symmetry, a small $d_c$ could be indicative of the presence of quark matter.} {In the right panel of Fig. \ref{fig:vs-density}, we illustrate the posterior distribution of the quantity $d_c$ as a function of density for all the cases under consideration, both including and excluding PREX-II data. In the right subfigure $(c)$, excluding PREX-II data, we display the confidence intervals for $d_c$ derived from {the NL, NL-$\sigma$ cut and NL-DM models}. The blue-shaded region represents the 90\% confidence interval for the NL model, with dashed blue lines indicating its boundaries. As density increases, $d_c$ initially decreases, followed by an increase, reaching a peak around 0.5~fm$^{-3}$, and subsequently exhibits a clear downward trend. For the NL-$\sigma$ cut model, depicted by the green hatched region with circle patterns and bounded by dashed green lines, a similar decreasing trend is observed, but it peaks earlier, around 0.4 fm$^{-3}$, and it is due to the stiffening of the $\sigma$ potential it. For the NL-DM model, the confidence region is narrower than those of the NL and NL-$\sigma$ cut models, and the peak is lower compared to the other models. The value of $d_c$ falls below 0.2 even though in the present approach, only nuclear matter exists at higher densities, specifically beyond 0.7~fm$^{-3}$. In contrast, the NL-$\sigma$ cut model shows a value below 0.2 at much lower densities, specifically beyond 0.6~fm$^{-3}$, even with only nuclear degrees of freedom. This behavior is due to the dominance of the repulsive $\omega$ potential, which is modified by the $\sigma$ potential. In the right subfigure ($d$), which incorporates PREX-II data, there is no discernible impact attributable to the inclusion of PREX-II data.}

Fig. \ref{fig:mass-radius-lam-DM} depicts the mass-radius (M-R)  and mass-tidal deformability (M-$\Lambda$) relationships of neutron stars based on the different scenarios considered in this study, compared with different astrophysical observational data. In panel (a), results for the NL (blue), NL-$\sigma$ cut (green with dots), and NL-DM (black dashed) models are plotted without including data from the PREX-II experiment. Panel (b) shows the same information, including the PREX-II data, in the likelihood, represented by the colors red, purple, and black, respectively. { In panels (c) and (d), we plot the 90\% CI of mass-tidal deformability for all the models with and without PREX-II, respectively.} {All distributions correspond to  90\% confidence intervals.} We compare our results with several observational constraints. In panels (a) and (b),  the grey region shows constraints from the binary components of the gravitational wave event GW170817, including their 90\% and 50\% credible intervals (CI). Constraints from the NICER x-ray data for the millisecond pulsar PSRJ0030+0451 are depicted in cyan and yellow color, while those for the pulsar PSRJ0740+6620 are shown in orange and peru color, both representing the 1$\sigma$ (68\%) CI for the 2D posterior distribution in the M-R domain. The new pulsar data PSR J0437-4715 is highlighted in a lilac shade.  In panels (c) and (d), the constraints obtained from GW170817 are included for the 1.36$M_\odot$ tidal deformability \cite{LIGOScientific:2018cki}.

In Fig. \ref{fig:mass-radius-lam-DM} panels (a) and (b),  the posterior distributions of the three different cases diverge from each other starting around an NS mass of 1.4 $M_\odot$. The NL-$\sigma$ cut tends to shift the M-R posterior to the right, thereby increasing the radius, while the dark matter in NL-DM tends to shift it to the left. Therefore, the effects of dark matter and the $\sigma$ cut potential are opposite. We aim to test which case is favored by the current astrophysical and nuclear constraints, {and therefore, we calculate the Bayes factor for each inference model. The results are given in Tables \ref{tab:bayes2} and  \ref{tab:bayes}, where, the Bayes factor for each model and the Bayes factor ratios are given, respectively.  Interestingly, our evidence calculations suggest that: i) the models that do not include the PREX-II constraints are favored, having systematically a higher Bayes factor; ii) considering the models with no PREX-II constraint,  the model with the $\sigma$ cut seems to be preferred with respect to both the NL and the NL-DM models, indicating a preference for a stiffening of the EOS at high densities. These results should be reflected on the slope of the M-R curves, giving preference for a smaller negative slope or even a positive slope.
}

\begin{table}[h!]
\setlength{\tabcolsep}{0.6pt}
\renewcommand{\arraystretch}{1.1}
    \centering
    \caption{Log evidence $\ln (\mathcal{Z})$ Values for the different Models.
    The {\it Best Model} is NL-$\sigma$ cut (without PREX-II)  with the highest log evidence of \(-62.18\).}
\begin{tabular}{lcc}
\hline \hline 
\textbf{Model}            & $\ln (\mathcal{Z})$       & \begin{tabular}[c]{@{}c@{}}$\ln (\mathcal{Z})$\\ (With  PSR J0437-471)\end{tabular} \\ \hline
NL                        & \(-64.14 \pm 0.16\)       & \( -65.25 \pm 0.15 \)                                                                \\
NL + PREX-II              & \(-68.53 \pm 0.17\)       &          ...                                                                           \\
                          &                           &                                                                                     \\
NL-$\sigma$ cut           & \({\it -62.18 \pm 0.15}\) & \( {\it -63.36 \pm 0.15} \)                                                                \\
NL-$\sigma$ cut + PREX-II & \(-66.15 \pm 0.17\)       &      ...                                                                               \\
                          &                           &                                                                                     \\
NL DM                     & \(-64.53 \pm 0.15\)       & \( -65.57 \pm 0.15 \)                                                                \\
NL DM + PREX-II           & \(-69.12 \pm 0.17\)       &         ...                                                                            \\ \hline
\end{tabular}
\label{tab:bayes2}
\end{table}

\begin{table}[h!]
\setlength{\tabcolsep}{1.pt}
\renewcommand{\arraystretch}{1.2}
    \centering
    \caption{Log Evidence Differences and Interpretations (P2 indicates with PREX-II, and NL-$\sigma$c indicates NL-$\sigma$ cut).}    
    \begin{tabular}{l r cl}
        \hline \hline
      \textbf{Model1/Model2}$\phantom{m}$ & \textbf{$\Delta \ln(\mathcal{Z})$} &$\phantom{m}$& \textbf{Interpretation} \\ \hline
     NL-$\sigma$c P2/NL-$\sigma$c  & \(-3.96\) &&{\footnotesize Decisive for NL-$\sigma$c}\\ %\hline
        NL-$\sigma$c P2/NL P2& \(2.38\) & & {\footnotesize  Substantial  for NL-$\sigma$c} P2\\ %\hline
        NL-$\sigma$c P2/NL & \(-2.01\) &&{\footnotesize  Substantial for  NL}\\ %\hline
        NL-$\sigma$c/NL P2 & \(6.35\)& &{\footnotesize  Decisive  for NL-$\sigma$c} \\ %\hline
        NL-$\sigma$c/NL & \(1.96\)& &{\footnotesize  Substantial  for NL-$\sigma$c} \\ %\hline
        NL P2/NL & \(-4.39\) && {\footnotesize Decisive for  NL} \\ \hline        
    \end{tabular}
 \label{tab:bayes}   
\end{table}

Fig. \ref{fig:dm_dr_noprex} depicts the dR/dM distribution for neutron stars with a mass of 1.6 $M_\odot$. The left panel shows the distribution across all three models without PREX-II data, while the right panel incorporates the PREX-II data. Among the models, the NL$-\sigma$cut is the most preferred, displaying a dR/dM slope of $-0.44^{+0.43}_{-0.82}$ for a neutron star mass of 1.6 $M_\odot$. The NL model has a slightly more negative slope of $-1.04_{-0.84}^{+0.53}$, and the NL-DM model yields the most negative one with a value  $-1.20_{-0.82}^{+0.52}$. When additional PREX-II data is considered, all slopes become more negative, and Bayesian evidence indicates lesser support for these models. The effect of PREX-II data makes the symmetry energy stiffer, increasing the radius for lower masses below 1.6 $M_\odot$. It can be seen from Fig. \ref{fig:mass-radius-lam-DM}(b) that the  PREX-II data makes the M-R distribution narrower in the lower part of the MR curve. This will be explained in greater detail at the end of the section.

\begin{table*}[]
\setlength{\tabcolsep}{1.5pt}
\renewcommand{\arraystretch}{1.5}
\caption{The median values and related 90\% confidence intervals (CI) for certain nuclear matter parameters (NMPs), equation of state (EOS), and neutron star (NS) properties for the following models: NL, NL-$\sigma$ cut, and NL-DM, both with and without the inclusion of PREX-II data. In this table, saturation density ($\rho_0$) is expressed in units of ${\rm fm}^{-3}$. NMPs such as $\epsilon_0$ - $Z_{\rm sym,0}$ are given in MeV. The properties of neutron stars, such as $M_{\rm max}$, are expressed in units of $M_{\odot}$. The radii corresponding to masses $M_i \in [1.4, 1.6, 1.8, 2.07]$ $M_\odot$ are given in kilometers. The parameter $\Lambda_{1.36}$, representing the tidal deformability, is unitless. The square of the speed of sound, $c_s^2$, is in units of $c^2$. The frequencies for the $f$ mode and the $p$ mode for different NS masses are measured in kHz.}
{\scriptsize \begin{tabular}{ccccccccccccc}
\hline \hline 
\multirow{3}{*}{\textbf{Quantity}}    & \multicolumn{6}{c}{Without PREX-II}                                                                                & \multicolumn{6}{c}{With PREX-II}                                                                                   \\ \cline{2-13} 
                                      & \multicolumn{2}{c}{NL}               & \multicolumn{2}{c}{NL-$\sigma$ cut}   & \multicolumn{2}{c}{NL DM}            & \multicolumn{2}{c}{NL}               & \multicolumn{2}{c}{NL-$\sigma$ cut}   & \multicolumn{2}{c}{NL DM}            \\
                                      & \textbf{Med.} & \textbf{CI}          & \textbf{Med.} & \textbf{CI}          & \textbf{Med.} & \textbf{CI}          & \textbf{Med.} & \textbf{CI}          & \textbf{Med.} & \textbf{CI}          & \textbf{Med.} & \textbf{CI}          \\ \hline
$m^*$             & 0.73          & {[}0.69, 0.78{]}     & 0.77          & {[}0.73, 0.79{]}     & 0.74          & {[}0.71, 0.78{]}     & 0.73          & {[}0.69, 0.78{]}     & 0.76          & {[}0.72, 0.78{]}     & 0.75          & {[}0.72, 0.78{]}     \\
$\rho_0$          & 0.160         & {[}0.155, 0.165{]}   & 0.160         & {[}0.155, 0.165{]}   & 0.161         & {[}0.155, 0.165{]}   & 0.160         & {[}0.155, 0.165{]}   & 0.160         & {[}0.155, 0.165{]}   & 0.161         & {[}0.156, 0.165{]}   \\
$\epsilon_0$      & -15.99        & {[}-16.03, -15.96{]} & -15.99        & {[}-16.03, -15.96{]} & -16.00        & {[}-16.03, -15.97{]} & -15.99        & {[}-16.03, -15.96{]} & -16.00        & {[}-16.03, -15.96{]} & -16.00        & {[}-16.03, -15.97{]} \\
$K_0$                                 & 239           & {[}215, 263{]}       & 240           & {[}222, 259{]}       & 243           & {[}227, 265{]}       & 238           & {[}212, 262{]}       & 236           & {[}216, 257{]}       & 244           & {[}230, 266{]}       \\
$Q_0$                                 & -472          & {[}-541, -401{]}     & -522          & {[}-592, -466{]}     & -467          & {[}-521, -404{]}     & -483          & {[}-563, -415{]}     & -537          & {[}-612, -468{]}     & -475          & {[}-524, -414{]}     \\
$Z_0$                                 & 2465          & {[}1406, 3718{]}     & 2060          & {[}1400, 2648{]}     & 2221          & {[}1436, 2973{]}     & 2395          & {[}1389, 3742{]}     & 2181          & {[}1445, 2977{]}     & 2119          & {[}1306, 2788{]}     \\
$J_{\rm sym,0}$    & 32            & {[}29, 36{]}         & 32            & {[}29, 36{]}         & 32            & {[}30, 36{]}         & 38            & {[}36, 39{]}         & 38            & {[}36, 39{]}         & 38            & {[}37, 40{]}         \\
$L_{\rm sym,0}$    & 54            & {[}38, 86{]}         & 54            & {[}38, 87{]}         & 54            & {[}37, 90{]}         & 106           & {[}97, 112{]}        & 104           & {[}96, 110{]}        & 105           & {[}97, 111{]}        \\
$K_{\rm sym,0}$    & -133          & {[}-175, -69{]}      & -158          & {[}-183, -87{]}      & -144          & {[}-176, -76{]}      & -11           & {[}-57, 18{]}        & -20           & {[}-70, 2{]}         & -17           & {[}-64, 8{]}         \\
$Q_{\rm sym,0}$    & 1004          & {[}50, 1294{]}       & 866           & {[}1, 1270{]}        & 965           & {[}2, 1297{]}        & 17            & {[}-48, 65{]}        & 8             & {[}-47, 49{]}        & 7             & {[}-54, 54{]}        \\
$Z_{\rm sym,0}$    & -2926         & {[}-10616, 1519{]}   & -1227         & {[}-8971, 1723{]}    & -2272         & {[}-10276, 1649{]}   & -600          & {[}-895, -22{]}      & -515          & {[}-704, 176{]}      & -567          & {[}-759, -2{]}       \\
$M_{\rm max}$      & 2.014         & {[}1.893, 2.152{]}   & 2.055         & {[}1.931, 2.163{]}   & 1.981         & {[}1.876, 2.095{]}   & 2.003         & {[}1.881, 2.131{]}   & 2.087         & {[}1.960, 2.201{]}   & 1.975         & {[}1.868, 2.079{]}   \\
$R_{\rm max}$      & 10.65         & {[}10.19, 11.02{]}   & 11.12         & {[}10.48, 11.47{]}   & 10.52         & {[}10.16, 10.82{]}   & 11.00         & {[}10.65, 11.31{]}   & 11.62         & {[}10.93, 11.96{]}   & 10.89         & {[}10.59, 11.15{]}   \\
$R_{1.4}$           & 12.18         & {[}11.77, 12.68{]}   & 12.29         & {[}11.81, 12.85{]}   & 12.09         & {[}11.74, 12.65{]}   & 13.15         & {[}12.84, 13.40{]}   & 13.20         & {[}12.83, 13.47{]}   & 13.10         & {[}12.79, 13.35{]}   \\
$R_{1.6}$ &	12.03 &	[11.57, 12.47]	& 12.25	& [11.68, 12.74] & 11.91 & [11.53, 12.38] & 12.82 & [12.43, 13.13] & 13.04 & [12.54, 13.37] & 12.74 & [12.37, 13.04] \\ 
$R_{1.8}$ &	11.75 &	[11.11, 12.24]	& 12.11	& [11.38, 12.58] & 11.59 & [11.07, 12.04] & 12.35 & [11.73, 12.79] & 12.81 & [12.09, 13.22] & 12.22 & [11.65, 12.64] \\ 
$R_{2.07}$          & 11.32         & {[}10.71, 11.88{]}   & 11.66         & {[}11.01, 12.25{]}   & 11.25        & {[}10.96, 11.57{]} & 11.65         & {[}11.05, 12.25{]}   & 12.29         & {[}11.51, 12.83{]}   & 11.66        & {[}11.29, 12.10{]} \\
$\Lambda_{1.36}$    & 440           & {[}356, 536{]}       & 478           & {[}363, 594{]}       & 419           & {[}351, 526{]}       & 653           & {[}554, 745{]}       & 675           & {[}548, 781{]}       & 639           & {[}541, 726{]}       \\
$c^2_{s, {\rm max}}$ & 0.56          & {[}0.48, 0.67{]}     & 0.47          & {[}0.39, 0.58{]}     & 0.58          & {[}0.49, 0.68{]}     & 0.53          & {[}0.46, 0.64{]}     & 0.44          & {[}0.38, 0.55{]}     & 0.56          & {[}0.48, 0.65{]}     \\
$f_{1.4}$ & 2.30 & {[}2.19, 2.39{]} & 2.27 & {[}2.15, 2.38{]} & 2.33 & {[}2.20, 2.41{]} & 2.10 & {[}2.04, 2.16{]} & 2.08 & {[}2.03, 2.17{]} & 2.11 & {[}2.06, 2.18{]} \\ 
$f_{1.6}$ & 2.38 & {[}2.29, 2.50{]} & 2.33 & {[}2.23, 2.47{]} & 2.42 & {[}2.32, 2.51{]} & 2.22 & {[}2.15, 2.31{]} & 2.17 & {[}2.10, 2.28{]} & 2.25 & {[}2.18, 2.33{]} \\ 
$f_{1.8}$ & 2.48 & {[}2.37, 2.64{]} & 2.39 & {[}2.29, 2.57{]} & 2.54 & {[}2.44, 2.66{]} & 2.36 & {[}2.26, 2.51{]} & 2.26 & {[}2.17, 2.42{]} & 2.41 & {[}2.32, 2.54{]} \\ 
$f_{2.0}$ & 2.56 & {[}2.44, 2.68{]} & 2.48 & {[}2.36, 2.69{]} & 2.65 & {[}2.55, 2.76{]} & 2.50 & {[}2.37, 2.69{]} & 2.36 & {[}2.25, 2.63{]} & 2.57 & {[}2.47, 2.71{]} \\ 
$p_{1.4}$           & 6.38          & {[}5.69, 6.94{]}   & 6.37          & {[}5.66, 6.97{]}   & 6.43          & {[}5.69, 6.99{]}   & 5.44          & {[}5.33, 5.57{]}   & 5.43          & {[}5.33, 5.57{]}   & 5.48          & {[}5.38, 5.61{]} \\ 
$p_{1.6}$           & 6.66          & {[}5.98, 7.11{]}   & 6.68          & {[}5.93, 7.22{]}   & 6.71          & {[}5.99, 7.16{]}   & 5.71          & {[}5.59, 5.87{]}   & 5.68          & {[}5.57, 5.84{]}   & 5.76          & {[}5.65, 5.91{]} \\ 
$p_{1.8}$           & 6.89          & {[}6.28, 7.23{]}   & 6.95          & {[}6.22, 7.38{]}   & 6.95          & {[}6.32, 7.28{]}   & 6.00          & {[}5.85, 6.23{]}   & 5.94          & {[}5.82, 6.15{]}   & 6.08          & {[}5.94, 6.28{]} \\ 
$p_{2.0}$           & 7.04          & {[}6.60, 7.25{]}   & 7.12          & {[}6.53, 7.41{]}   & 7.11          & {[}6.72, 7.31{]}   & 6.30          & {[}6.11, 6.58{]}   & 6.23          & {[}6.07, 6.60{]}   & 6.41          & {[}6.23, 6.62{]}   \\ \hline
\end{tabular} }
\label{tab_3}
\end{table*}
 
In Table \ref{tab_3}, we summarize some properties of nuclear matter properties (NMP) and neutron star properties predicted by the NL, NL-$\sigma$ cut, and NL-DM inference models. {Some conclusions can be drawn: i) independently of the inclusion or not of PREX-II constraints the symmetric nuclear matter properties are only slightly affected by the inclusion of the $\sigma$ cut potential and the presence of DM, in particular, in the last case making the EOS slightly harder to compensate the effect of  DM on the neutron star properties; ii) the symmetry energy and its slope at saturation do not depend on the model but the inclusion of PREX-II constraint rises the symmetry energy and its slope at saturation from 32 MeV and 54 MeV to 38 MeV and $\sim 105$ MeV, respectively.}

In  Fig. \ref{fig:mass-radius-lam-DM}(a) (no  PREX-II data), the NL model's confidence interval (shaded blue) shows a broader range of radii for neutron stars, predicting a median value of 12.18 km for a 1.4$M_{\odot}$ star, $R_{1.4}$,  see Table \ref{tab_3}. For the NL-$\sigma$ cut model, $R_{1.4}$ slightly increases to 12.29 km,  and for NL-DM it slightly decreases to 12.09 km. Including  PREX-II data, the radius shifts to larger values, in particular, to 13.15 km, 13.20 km, and 13.10 km for, respectively,  NL, NL-$\sigma$ cut, and NL-DM models. Including $\sigma$ cut increases the maximum mass from 2.055 $M_{\odot}$ to 2.087 $M_{\odot}$ with PREX-II. Conversely, adding DM decreases the maximum mass to 1.981 $M_\odot$ (or  1.975 $M_\odot$ with PREX-II). 

In Fig., \ref{fig:mass-radius-lam-DM}(c) and (d) showcase the 90\% confidence intervals for tidal deformability $\Lambda$, respectively without and with PREX-II constraint, as it relates to neutron star mass across all considered cases. Including PREX-II results in an increase in tidal deformability for all cases due to an increase in radius. The effect of the  $\sigma$ cut parameter and DM matter on the tidal deformability is clearly seen comparing $\Lambda_{1.36} = 478$ for the NL-$\sigma$ cut model and $\Lambda_{1.36} =419$ for NL-DM with $\Lambda_{1.36} = 440$ for the NL model,  (see Table \ref{tab_3}). PREX-II data shifts $\Lambda_{1.36}$ predictions to higher values: $\Lambda_{1.36} = 653$ for NL-PREX-II (red), $\Lambda_{1.36} = 675$ for NL-$\sigma$ cut-PREX-II (purple), and $\Lambda_{1.36} = 639$ for NL-DM-PREX-II (black boundary). Both panels (c) and (d)  show that all models fit within or near GW170817 constraints, even when  PREX-II data are included.

\begin{figure}
    \centering
    \includegraphics[width=0.4\textwidth]{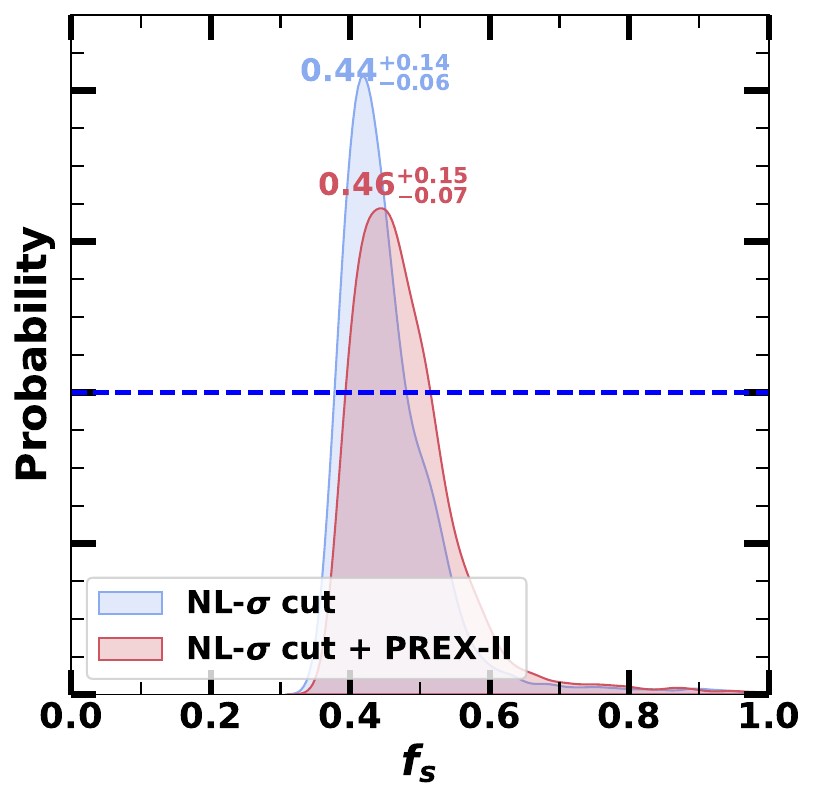}\\
        \includegraphics[width=0.4\textwidth]{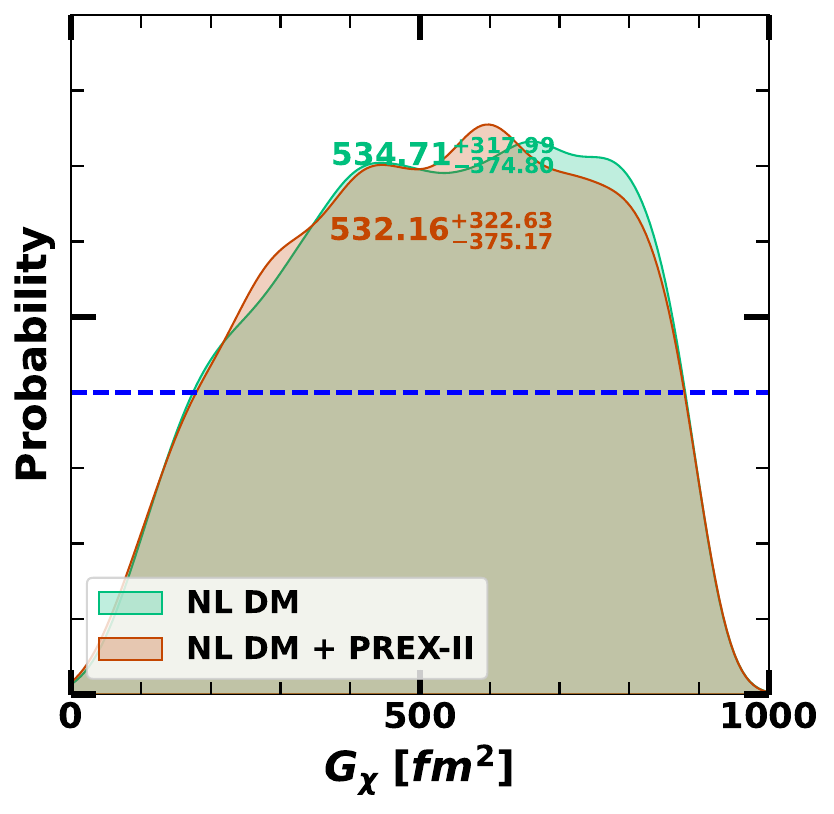}
    \caption{The probability distribution of the free parameter $f_s$ for the $\sigma$cut model (top) and  $G_{\chi}$ for the DM model (bottom)  is shown. Both panels include the distribution with and without PREX-II. The blue line indicates uniform prior.}
    \label{fig:fs}
\end{figure}

Given that the inference analysis has been completed, we can now explore the constraints on the unknown parameter $f_s$ for the NL-$\sigma$ cut and the $G_\chi$ parameter for the dark matter model, under the assumption of a uniform prior for both cases. The values considered for $f_s$ ranged from 0 to 1, and for $G_\chi$, from 0 to 1000 (fm$^2$). The justification for $G_\chi$ prior range can be found in Ref. \cite{Shirke:2023ktu}. 
Figure \ref{fig:fs} shows the probability distribution of $f_s$ for the $\sigma$cut model  (top panel) and of  $G_{\chi}$ for the DM  model from the  Bayesian inference. In both panels, results including and without PREX-II data are shown. Including PREX-II data shifts the median value of $f_s$ slightly from 0.438 to 0.464 and broadens the distribution.
Figure \ref{fig:fs} bottom panel shows that $G_{\chi}$ is only marginally lowered by  PREX-II data, suggesting a minor shift in the parameter estimate. However, the spread in $G_{\chi}$ remains relatively unchanged. We conclude that the wide range of constraints from the nuclear and astrophysical data considered does not significantly narrow down this parameter.

\begin{figure}
    \centering
    \includegraphics[width=0.50\textwidth]{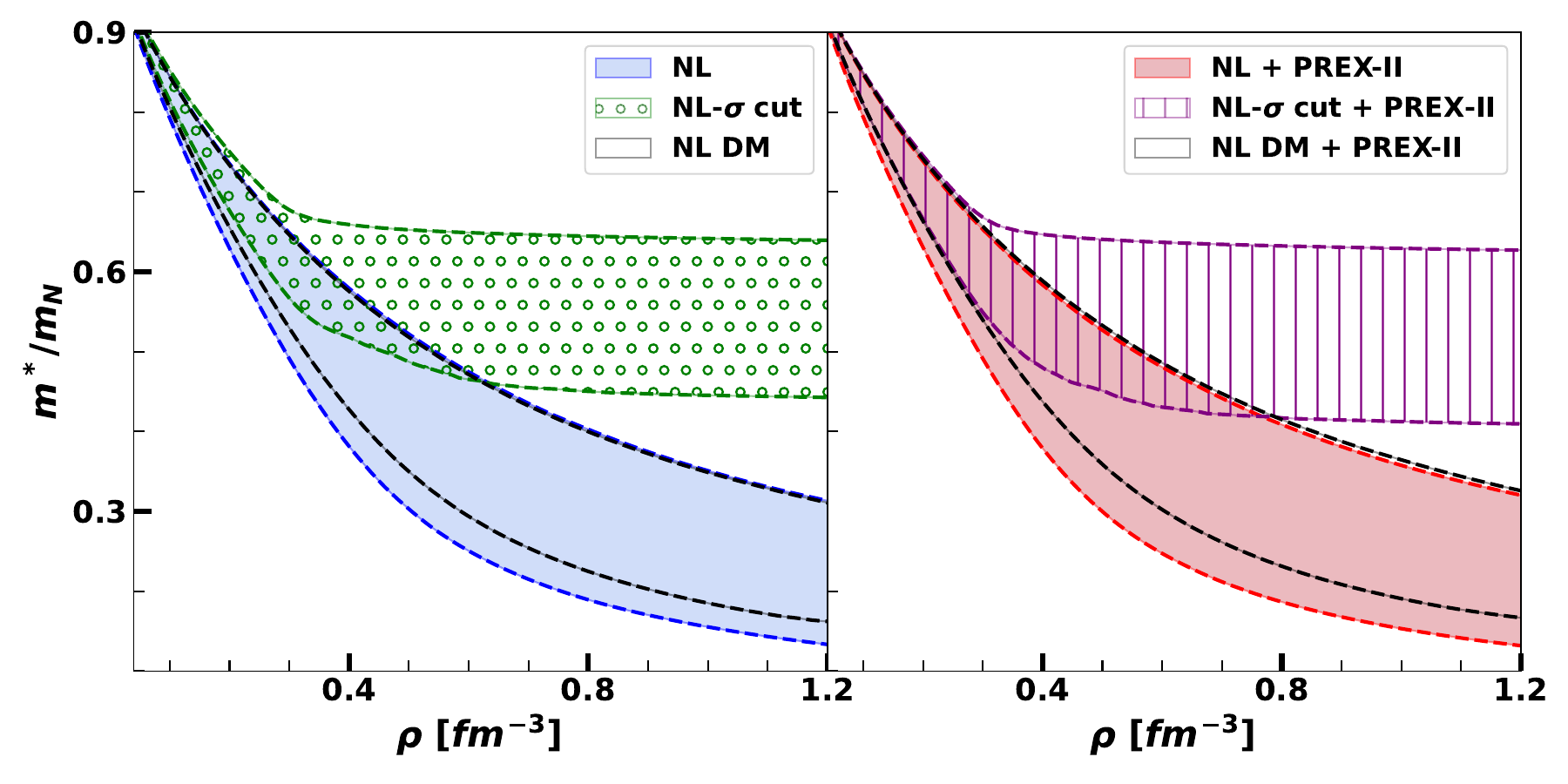}
    \caption{The 90\% CI of effective mass (\(m^*\)) as a function of baryon density (\(\rho\)) for the NL, NL-\(\sigma\)-cut and NL-DM cases. The left panel with all the constraints in Table \ref{tab1} except for the PREX-II data, while the right panel includes the PREX-II data.}
    \label{fig:m_dkl}
\end{figure}

Fig. \ref{fig:m_dkl} depicts the 90\% CI region for the effective mass $m^*$ against baryon density $\rho$, without (left panel) and with (right panel) PREX-II constraints. The NL scenario demonstrates a decreasing trend with density, and PREX-II data exerts only a minimal effect on $m^\star$. When the $\sigma$ cut potential is applied, the effective mass stabilizes above a 0.3 fm$^{-3}$ density due to the  $\sigma$ potential, thereby stiffening the EOS. For the NL-DM model that includes admixed dark matter, there is a slight upward pull on the $m^{\star}$ posterior, but it remains encompassed by the NL model in both panels. 

\begin{figure}
    \centering
    \includegraphics[width=0.5\textwidth]{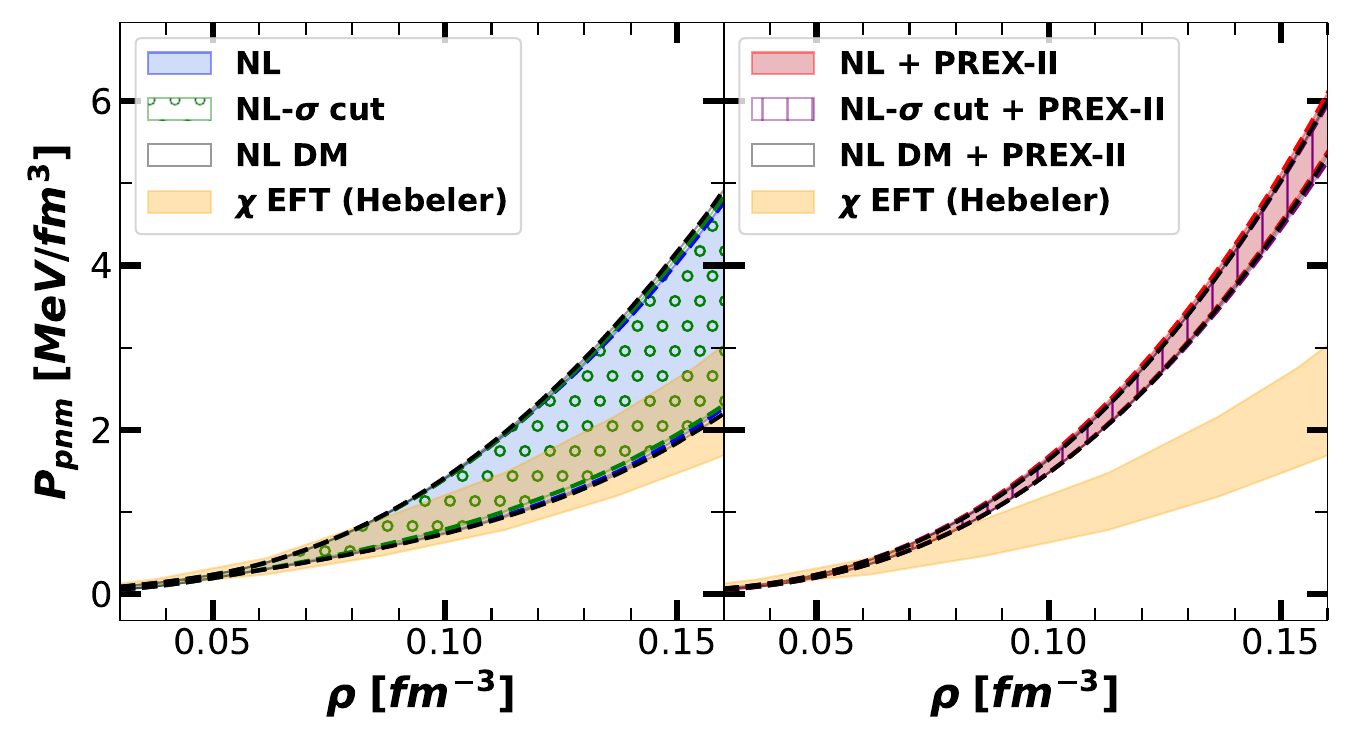}
    \caption{Displayed are the posterior distributions for the NL, NL-$\sigma$ cut, and NL-DM models within the 90\% confidence interval (CI) regarding the pressure (P) as a function of baryon density in pure neutron matter (PNM). The plot on the left does not include PREX-II, while the plot on the right does.}
    \label{fig:pnm-density}
\end{figure}
In Fig. \ref{fig:pnm-density}, we evaluate how well the computed posteriors match the {\it ab-initio} chiral effective field theoretical ($\chi$EFT) calculations for pure neutron matter (PNM) constraints across all scenarios. The figure displays the posterior distributions for the NL and NL-$\sigma$ cut models within the 90\% confidence interval (CI) for the PNM pressure as a function of baryon density. At low densities near saturation density, all three cases, both with and without PREX-II, are indistinguishable.  At larger densities, the inclusion of additional PREX-II data makes the pure neutron matter pressure significantly stiffer, pulling the posterior outside the PNM chEFT constraints. In the absence of  PREX-II constraint, the PNM pressure posterior for all three cases shows a good overlap with $\chi$EFT data. It should be noted that the $\chi$EFT PNM constraints were not part of the constraints applied to the likelihood considered in this study. {This data indicates the PREX-II data are in tension with $\chi$EFT constraints, contrary to all the other constraints included in our inference analysis, independently of the model considered.}

%\section{Non-radial Oscillations}
\begin{figure}
    \centering
    \includegraphics[width=0.50\textwidth]{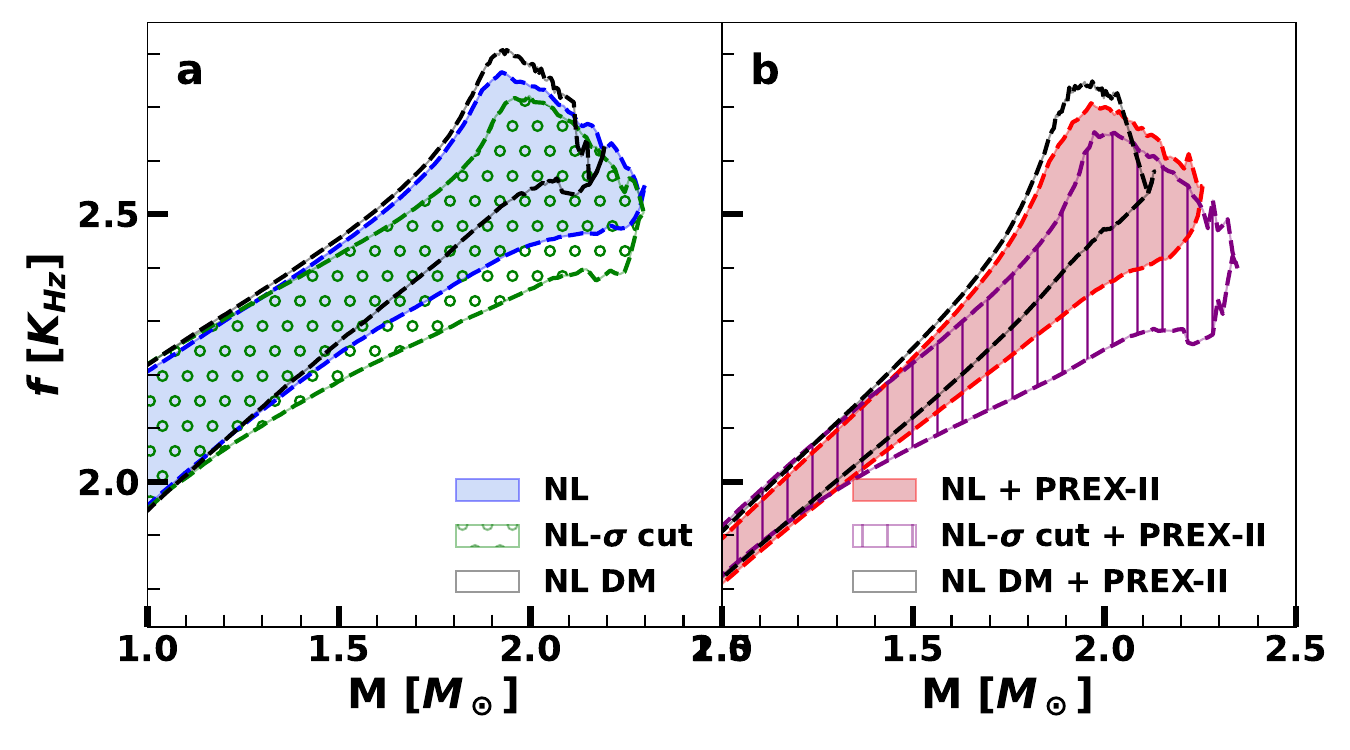}
    \includegraphics[width=0.50\textwidth]{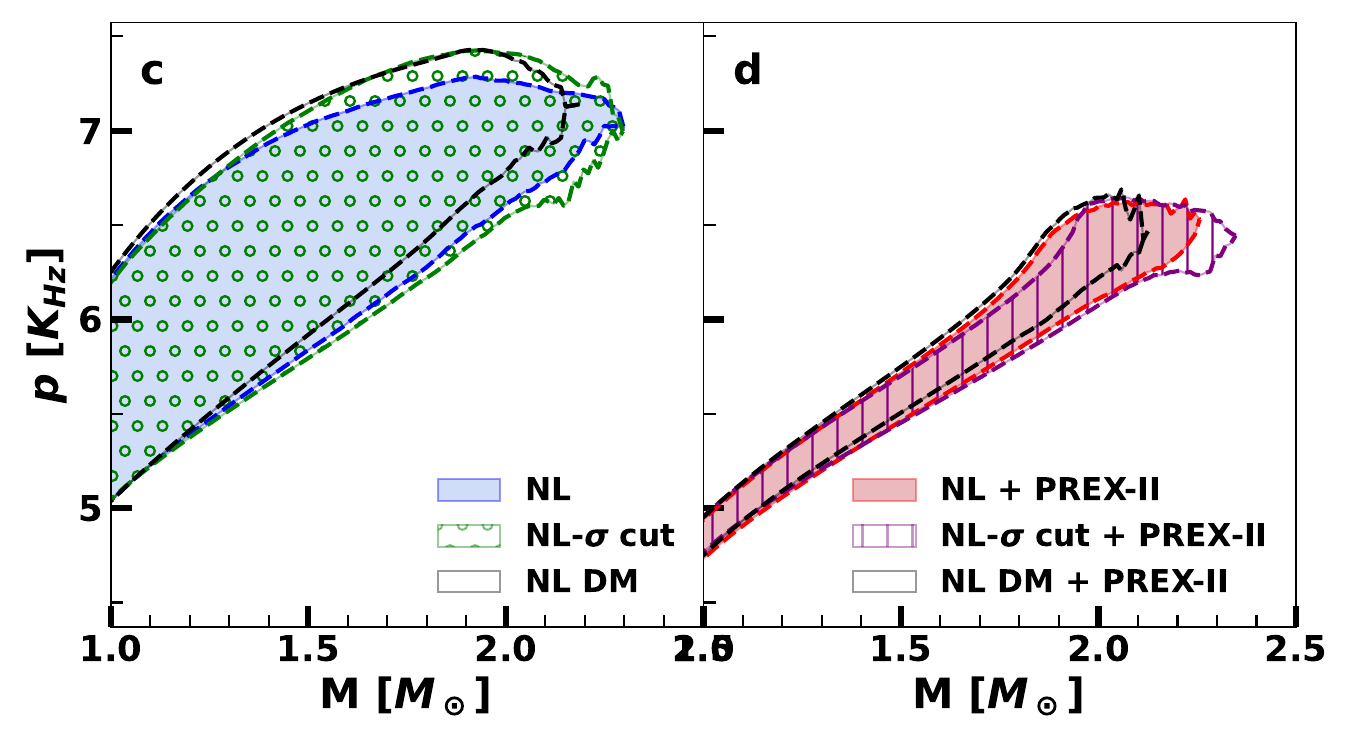}
    \caption{The graphs illustrate the relationship between neutron star (NS) mass (\(M\) in units of solar mass \(M_\odot\)) and the frequencies of non-radial oscillation modes: \(f\) (fundamental mode, upper plots) and \(p\) (pressure mode, lower plots). The left panels show results for the NL, NL \(\sigma\)-cut, and NL-DM with dark matter, while the right panels compare the same but with additional PREX-II data. The domain represents the 90\% CI region.}
    \label{fig:f_dm_mode}
\end{figure}

\begin{figure*}
    \centering
    \includegraphics[width=1.0\textwidth]{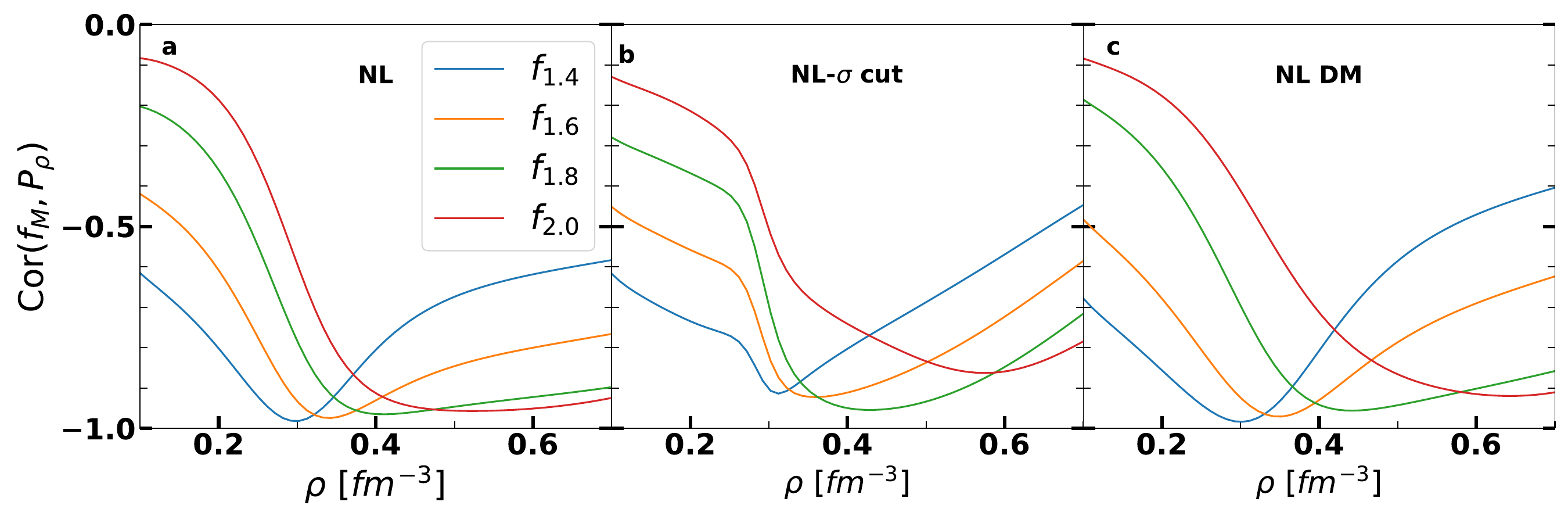}
    \caption{Pearson's correlation coefficients between the $f$ mode oscillation frequency $f_{M}$ for neutron star masses $M\in[1.4,1.6,1.8,2.0]$ $M_\odot$ and the neutron star matter pressure $P(\rho)$ across different baryon densities $\rho$, for the models without PREX-II.}
    \label{fig:f_mode_co}
\end{figure*}

\begin{figure}
    \centering
    \includegraphics[width=0.4\textwidth]{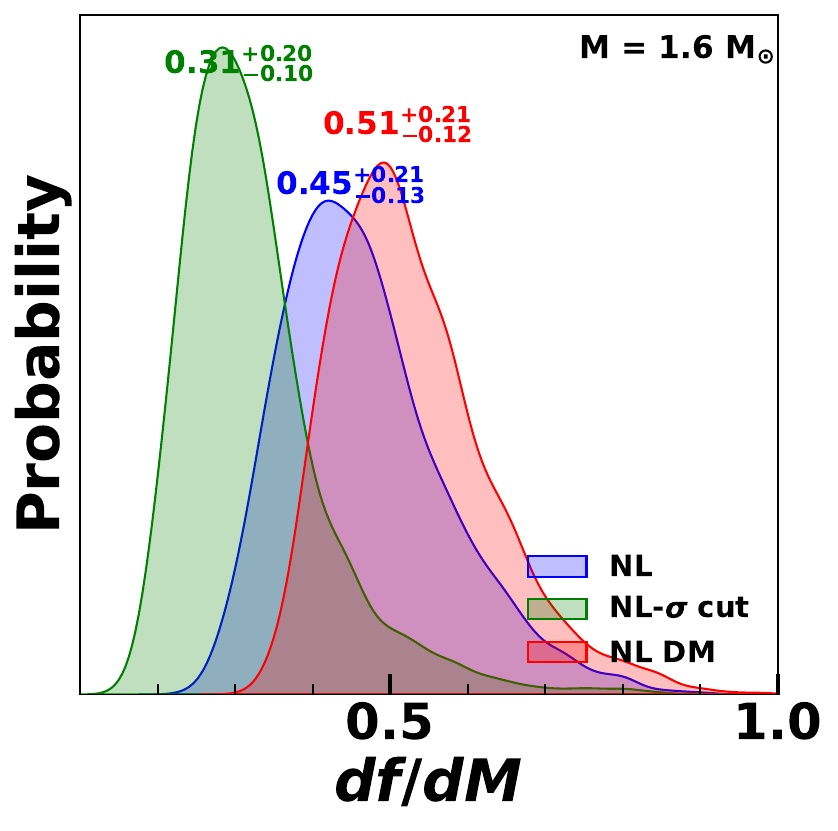}
    \caption{ The $df/dM$ distribution at a neutron star mass of 1.6 $M_\odot$ for three scenarios: NL, NL$\sigma$cut, and NL-DM, without PREX-II data.}
    \label{fig:df_dm_prex}
\end{figure}

We have computed the non-radial oscillation modes of neutron stars, specifically the frequencies of the $f$ and $p$ modes, within the Cowling approximation using the posterior results of the three cases: Nl, NL-$\sigma$ cut, and NL-DM. Figure \ref{fig:f_dm_mode} illustrates the $f$-mode (fundamental) and $p$-mode (pressure) frequencies of neutron stars as functions of their mass (in solar masses). The right column displays corresponding data incorporating PREX-II results. The $f$-mode frequency ranges approximately from 2.0 kHz to 2.8 kHz as the neutron star mass increases from 1.0 to 2.3 solar masses. The $p$-mode frequency spans from 5 kHz to 7.5 kHz over the same mass range. For PREX-II, the distribution is narrower at lower neutron star masses, an effect similar to the one obtained in the mass-radius plot. The feature $dM/df$ varies among the three cases for the $f$-mode frequency. In contrast, the $p$-mode frequency is only marginally distinguishable across all three cases and columns. The $dM/dp$  slope for the $p$-mode and mass domain is similar for all three scenarios. Compared to the $p$ mode, the $f$-mode is more sensitive to the symmetry energy and conveys analogous slope information as the mass-radius relationship. This correlation is evident as   Ref \cite{Roy:2023gzi} established a strong relationship between NS radius and $f$-mode frequency. 

Fig. \ref{fig:f_mode_co} displays the Pearson correlation coefficients between pressure and $f$-mode frequencies for neutron stars of different masses. These coefficients are plotted against the baryon number density ($\rho_B$). The figure comprises three panels corresponding to the models NL (left), NL-$\sigma$ cut (middle), and NL-DM (right, all not including the PREX-II constraint.  The different colored lines correspond to neutron stars of varying masses, specifically $1.4 M_{\odot}$ (blue), $1.6 M_{\odot}$ (orange), $1.8 M_{\odot}$ (green), and $2.0 M_{\odot}$ (red). The correlation peaks at different densities for the $f$ mode, depending on the mass. As the mass increases from 1.4 to 2 $M_\odot$, the correlation peak shifts to a higher density, which is consistent with the expectation that central density for a higher mass star resides at a higher density.  The correlations are stronger in the NL model, although they are all comparable: for the NL-$\sigma$ cut model, the onset of the $\sigma$cut potential seems to reduce the correlation associated with the lower mass stars. This potential also affects the correlation in the most massive stars; for the  NL-DM model, the correlation is more affected for the massive stars when the effect of DM is stronger.
 For higher NS masses, the correlation is weaker compared to others, in particular, for NL-DM and NL-$\sigma$ cut models. 

{In Fig. \ref{fig:df_dm_prex}, we plot the   $df/dM$ distribution at a neutron star mass of 1.6 $M_\odot$ for three scenarios: NL, NL$\sigma$cut, and NL-DM, without PREX-II data. The slope $df/dM$ is positive in all these cases, in contrast to the slope $dR/dM$ (see Fig. \ref{fig:dm_dr_noprex}). Moreover, it becomes apparent that there exists an inverse correlation between the neutron star (NS) radius and the $f$-mode oscillation frequency \cite{Kumar:2023rut,Roy:2023gzi}. The slope $df/dM$ for NL, NL-$\sigma$ cut, and NL-DM are $0.45^{+0.21}_{-0.13}$, $0.31^{+0.20}_{-0.10}$, and $0.51^{+0.21}_{-0.12}$, respectively. Including dark matter in the NL-DM model results in the steepest slope in the mass-frequency plane for $f$ mode oscillation. Given that the NL-$\sigma$ cut case exhibits the strongest Bayesian evidence compared to others (see Table \ref{tab:bayes2}), we can conclude that the collective data from nuclear to astrophysical sources favors a smaller slope $df/dM$, indicative of a stiff EOS.}

\subsection*{Impact of PSR J0437-4715}
\begin{figure*}
    \centering
    \includegraphics[width=1.0\textwidth]{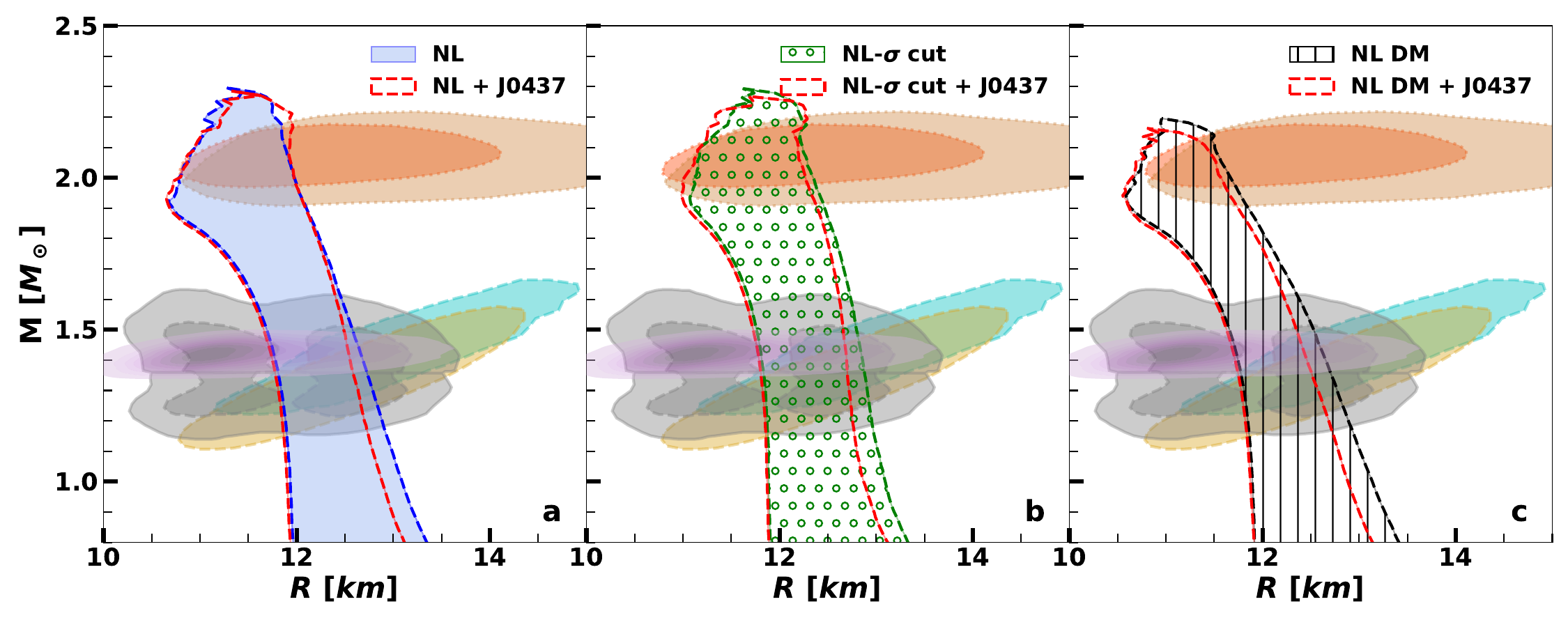}
    \caption{The posterior distribution of the neutron star mass-radius $P(R|M)$ for the models (a) NL, (b) NL-\(\sigma\)cut, and (c) NL-DM is compared with the distribution that includes the new PSR J0437-4715 NICER mass-radius measurements. The legends NL+J0437, NL-\(\sigma\)cut+J0437, and NL-DM+J0437 indicate the integration of PSR J0437-4715 NICER data with the previous NICER measurements. The additional constraints being compared are identical to those in Fig. \ref{fig:mass-radius-lam-DM}. Refer to its caption for more information.}
    \label{fig:dm_jo437}
\end{figure*}

NICER's nearest and brightest target is the 174 Hz millisecond pulsar PSR J0437-4715. Using NICER data from July 2017 to July 2021, and incorporating NICER background estimates, the latest mass-radius measurements of PSR J0437-4715 are reported \cite{Choudhury:2024xbk}. We have investigated the effect of these measurements, together with the two old NICER measurements for pulsars PSR J0030+0451 and PSR J0740+6620, on NL, NL-$\sigma$ cut, and NL-DM. All were compared without PREX-II.  Fig. \ref{fig:dm_jo437} shows the posterior distribution of neutron star mass-radius relations for the NL, NL-$\sigma$ cut, and NL-DM models, incorporating the new data from PSR J0437-4715 by NICER. Panels (a), (b), and (c) represent NL, NL-$\sigma$ cut, and NL-DM, respectively, and illustrate the effect of the new NICER measurements on the mass-radius posterior.  The inclusion of data from PSR J0437-4715 particularly affects the estimated radius for neutron stars in the 1 to 1.5 M$_\odot$ mass range. This new data reduces the upper limit of the 90\% confidence interval by about 200~m and the lower limit by less than $\sim 30$~m, with a consistent effect across all models, see Table \ref{tab5} where the NS radius for masses of 1.2, 1.4, and 1.6 M$_\odot$ for the three cases, including the new PSR constraints are given. {We have computed the Bayes evidence for each model that incorporates PSR J0437-4715 data but excludes PREX-II and observed a decrease of $\sim$ 1 in the logarithm of the Bayes evidence in all instances, see Table \ref{tab:bayes2}. This suggests that the new NICER data conflicts with the old data or that the current EOS model lacks the flexibility to simultaneously accommodate all NICER data.}

\begin{table*}[]
\caption{The median and 90\% confidence intervals (CI) of the radius \(R_M\) (km) for neutron star masses \(M \in [1.2, 1.4, 1.6]\) $M_\odot$, comparing results for obtained posterior with older NICER data and the additional inclusion of new PSR J0437-4715 NICER mass-radius measurements. The models include NL, NL-\(\sigma\) cut, and NL-DM. The terms NL+J0437, NL-\(\sigma\)cut+J0437, and NL-DM+J0437 represent the inclusion of PSR J0437-4715 NICER measurements along with the older NICER data. \label{tab5}}
\setlength{\tabcolsep}{2.5pt}
\renewcommand{\arraystretch}{1.5}
{\scriptsize \begin{tabular}{ccccclcccclcccc}
\hline \hline 
\multirow{2}{*}{\textbf{Quantity}} & \multicolumn{2}{c}{NL}             & \multicolumn{2}{c}{NL +J0437}      &  & \multicolumn{2}{c}{NL-$\sigma$ cut} & \multicolumn{2}{c}{NL-$\sigma$ cut + J0437} &  & \multicolumn{2}{c}{NL DM}          & \multicolumn{2}{c}{NL DM + J0437}  \\ \cline{2-5} \cline{7-10} \cline{12-15} 
                                   & \textbf{Med.} & \textbf{90\% CI}   & \textbf{Med.} & \textbf{90\% CI}   &  & \textbf{Med.}  & \textbf{90\% CI}   & \textbf{Med.}      & \textbf{90\% CI}       &  & \textbf{Med.} & \textbf{90\% CI}   & \textbf{Med.} & \textbf{90\% CI}   \\ \cline{2-15} 
$R_{1.2}$                          & 12.27         & {[}11.89, 12.89{]} & 12.23         & {[}11.85, 12.70{]} &  & 12.29          & {[}11.87, 12.95{]} & 12.21              & {[}11.85, 12.76{]}     &  & 12.20         & {[}11.85, 12.89{]} & 12.14         & {[}11.82, 12.69{]} \\
$R_{1.4}$                          & 12.18         & {[}11.78, 12.69{]} & 12.14         & {[}11.75, 12.55{]} &  & 12.29          & {[}11.81, 12.86{]} & 12.21              & {[}11.79, 12.69{]}     &  & 12.09         & {[}11.75, 12.65{]} & 12.03         & {[}11.72, 12.46{]} \\
$R_{1.6}$                          & 12.03         & {[}11.57, 12.47{]} & 11.98         & {[}11.54, 12.41{]} &  & 12.25          & {[}11.68, 12.74{]} & 12.17              & {[}11.64, 12.62{]}     &  & 11.91         & {[}11.53, 12.38{]} & 11.85         & {[}11.50, 12.23{]} \\ \hline
\end{tabular}}
\end{table*}

\begin{table*}[]
\setlength{\tabcolsep}{1.2pt}
\renewcommand{\arraystretch}{1.5}
\caption{The median and 90\% confidence interval (CI) values for the derived parameters across various posteriors. Specifically, $B$ and $C$ are $b \times 10^3$ and $c \times 10^3$, respectively. The parameter $f_{s}$ in the NL - $\sigma$ cut model is dimensionless, while the parameter $G_{\chi}$ in the NL DM model is measured in units of ${\rm fm}^2$.}
{\scriptsize \begin{tabular}{ccccccccccccc}
\hline \hline 
\multirow{3}{*}{\textbf{Quantity}} & \multicolumn{6}{c}{Without PREX-II}                                                                              & \multicolumn{6}{c}{With PREX-II}                                                                                \\ \cline{2-13} 
                                   & \multicolumn{2}{c}{NL}              & \multicolumn{2}{c}{NL-$\sigma$ cut} & \multicolumn{2}{c}{NL DM}            & \multicolumn{2}{c}{NL}              & \multicolumn{2}{c}{NL-$\sigma$ cut} & \multicolumn{2}{c}{NL DM}           \\
                                   & \textbf{Med.} & \textbf{CI}         & \textbf{Med.} & \textbf{CI}         & \textbf{Med.} & \textbf{CI}          & \textbf{Med.} & \textbf{CI}         & \textbf{Med.} & \textbf{CI}         & \textbf{Med.} & \textbf{CI}         \\ \hline
$g_{\sigma}$                       & 8.438         & {[}7.827, 8.915{]}  & 7.965         & {[}7.681, 8.470{]}  & 8.257         & {[}7.842, 8.653{]}   & 8.369         & {[}7.775, 8.907{]}  & 8.073         & {[}7.714, 8.576{]}  & 8.144         & {[}7.753, 8.552{]}  \\
$g_{\omega}$                       & 9.903         & {[}8.689, 10.715{]} & 8.901         & {[}8.381, 9.927{]}  & 9.540         & {[}8.719, 10.242{]}  & 9.767         & {[}8.612, 10.687{]} & 9.120         & {[}8.475, 10.121{]} & 9.336         & {[}8.557, 10.082{]} \\
$g_{\rho}$                         & 10.108        & {[}9.282, 10.950{]} & 10.059        & {[}9.311, 10.840{]} & 10.097        & {[}9.274, 10.927{]}  & 9.288         & {[}9.048, 9.562{]}  & 9.364         & {[}9.142, 9.625{]}  & 9.342         & {[}9.118, 9.586{]}  \\
B                                  & 5.199         & {[}4.167, 7.897{]}  & 7.253         & {[}4.961, 8.731{]}  & 5.746         & {[}4.578, 7.880{]}   & 5.346         & {[}4.172, 8.125{]}  & 6.722         & {[}4.678, 8.612{]}  & 6.066         & {[}4.767, 8.257{]}  \\
C                                  & -4.106        & {[}-4.916, 0.346{]} & -2.124        & {[}-4.692, 3.478{]} & -3.838        & {[}-4.901, -0.032{]} & -3.969        & {[}-4.897, 1.540{]} & -2.613        & {[}-4.758, 2.709{]} & -3.441        & {[}-4.861, 1.784{]} \\
$\xi$                             & 0.005         & {[}0.000, 0.012{]}  & 0.011         & {[}0.001, 0.028{]}  & 0.004         & {[}0.000, 0.012{]}   & 0.006         & {[}0.000, 0.015{]}  & 0.015         & {[}0.002, 0.032{]}  & 0.005         & {[}0.000, 0.013{]} 
\\
$\Lambda_\omega$                          & 0.047         & {[}0.013, 0.092{]}  & 0.060         & {[}0.014, 0.104{]}  & 0.051         & {[}0.012, 0.096{]}   & 0.001         & {[}0.000, 0.006{]}  & 0.002         & {[}0.000, 0.008{]}  & 0.002         & {[}0.000, 0.007{]}  \\ 
$f_{s}$                         & -             & -                   & 0.44          & {[}0.38, 0.58{]}    & -             & -                    & -             & -                   & 0.46          & {[}0.39, 0.61{]}    & -             & -                   \\
$G_{\chi}$                         & -             & -                   & -             & -                   & 534.71        & {[}159.91, 852.7{]}  & -             & -                   & -             & -                   & 532.16        & {[}156.99, 854.79{]}  \\ \hline
\end{tabular}}
\end{table*}

\section{Conclusion \label{conclussion}}
In this study, we have examined the equation of state (EOS) for neutron stars under three distinct scenarios: a purely nucleonic composition, a nucleonic composition with a $\sigma$-cut potential, and a nucleonic composition with an admixture of dark matter.  The effect of the $\sigma$-cut potential is to stiffen the EOS above saturation density, having a net effect similar to the presence of a quarkyonic phase, see \cite{McLerran:2018hbz}, or an exclusion volume \cite{Typel:2016srf}. By employing Bayesian inference and incorporating the latest constraints from nuclear physics and astrophysical observations, we have been able to evaluate the plausibility and impact of each scenario. Our analysis reveals that the inclusion of dark matter and modified potentials in the EOS significantly affects the macroscopic properties of neutron stars, such as their mass-radius relationships and non-radial oscillation modes.

{Our results indicate that the inclusion of PREX-II constraints has a strong effect on several NS properties, such as mass-radius curves or $f$-modes, and in particular, on the slope of these curves. PREX-II data shifts the radius of low-mass stars to very large radii of the order of 13.5 - 14 km. The models including PREX-II data {entirely unsuccessful} to reproduce $\chi$EFT PNM pressure. Also, the calculation of the Bayes factor has shown decisive or substantial evidence against these models when compared with the models, not including PREX-II data.}

{The analysis of the effect of the $\sigma$ cut potential has shown that the constraints imposed in our Bayesian inference calculation favor this model, giving larger Bayes factors.  The NL-$\sigma$ cut model gives rise to a stiffening of the EOS at large densities and, therefore, predicts massive stars with larger radii and smaller $f$ mode frequencies. It also presents a very distinctive effect on the speed of sound, giving rise to a steep increase above 0.2 fm$^{-3}$ and a leveling out above 0.4 fm$^{-3}$. Also, the trace anomaly-related quantity was affected,  showing a clear peak for $\rho\sim 0.3$ fm$^{-3}$ followed by a steep decrease attaining values below 0.2 at 0.6 fm$^{-3}$ while for the other models, there is no distinctive peak and the values 0.2 is only reached above 0.8 fm$^{-3}$.}

{Moreover, our investigation also examined the non-radial oscillations, specifically the $f$ and $p$ modes. We identified a large sensitivity of the $f$ oscillations to changes in the neutron star's composition and EOS. Although working within the Cowling approximation, which in future work should be generalized to incorporate full general relativistic effects,  it was shown the existence of a strong correlation between the $f$-modes and the NS pressure at different densities, with the correlation peak shifting to a higher density as the mass increases. We also analyzed the slope of the $f$ mode curve with respect to the star mass. The smallest one was associated with the NL-$\sigma$ cut model, the model that presented the largest values of $dR/dM$, possibly even positive, due to the presence of a stiff  EOS at high densities. This is the most favored model.
}

{We have also investigated the impact of the new PSR J0437-4715 measurements on the neutron star mass-radius posterior distribution, observing a consistent reduction of approximately 0.2 km in the upper boundary of the 90\% confidence interval across all models. This refinement enhances the model fit, as evidenced by the notable decrease in the logarithmic Bayes evidence ($\sim 1$), suggesting either a conflict with previous measurements or a need for more flexible theoretical models to accommodate the updated data.}

\begin{acknowledgments}
This research is part of the Advanced Computing Project 2024.14108.CPCA.A3,  RNCA (Rede Nacional de Computação Avançada), funded by the FCT (Fundação para a Ciência e a Tecnologia, IPP, Portugal), and this work was produced with the support of Deucalion HPC, Portugal. The author, P.T, would like to acknowledge CFisUC, University of Coimbra, for their hospitality and local support during his visit in May - July 2023 for the purpose of conducting part of this research. The author T.M and C.P also acknowledges the support through national funds from FCT for the projects UIDB/04564/2020 and UIDP/04564/2020, identified by DOIs 10.54499/UIDB/04564/2020 and 10.54499/UIDP/04564/2020, respectively, and for project 2022.06460.PTDC with the DOI identifier 10.54499/2022.06460.PTDC. A. D. acknowledges the New Faculty Seed Grant (NFSG), NFSG/PIL/2024/P3825, provided by the Birla Institute of Technology and Science, Pilani, India.
\end{acknowledgments}

%\newpage
%\bibliography{apssamp}
\end{document}